\documentclass[twocolumn,showpacs,preprintnumbers,amsmath,amssymb]{revtex4}
\usepackage[english]{babel}

\usepackage{graphicx}

\newcommand{\beq}{\begin{equation}}
\newcommand{\eeq}{\end{equation}}
\newcommand{\bma}{\begin{math}}
\newcommand{\ema}{\end{math}}
\newcommand{\beqa}{\begin{eqnarray}}
\newcommand{\eeqa}{\end{eqnarray}}
\newcommand {\ee} {\end{eqnarray}}

\def\half{\frac{1}{2}}
\def\opone{\le\textbf{}\textbf{}avevmode\hbox{\small1\kern-3.8pt\normalsize1}}

\newcommand{\pref}[1]{(\ref{#1})}

\newcommand{\av}[1]{\langle #1\rangle}

\newcommand{\rmd}{{\mathrm d}}

\newcommand{\mbf}{\mathbf}

\newcommand{\zb}{\bar z}
\newcommand{\wb}{\bar w}
\newcommand{\imag}{\mathrm{Im}}

\newcommand{\vect}[1]{\mathbf #1}

\newcommand\ie {{\it i.e. }}
\newcommand\eg {{\it e.g. }}

\newcommand{\be}[1]{ \begin{eqnarray} \mbox{$\label{#1}$} }

\begin{document}

\title {Quantum Hall wave functions on the torus}

\author{M. Hermanns$^1$}
\author{J. Suorsa$^2$}
\author{E.J. Bergholtz$^1$}
\author{T.H. Hansson$^1$}
\author{A. Karlhede$^1$}

\affiliation{$^1$ Department of Physics, Stockholm University,\\ AlbaNova University Center, SE-106 91 Stockholm,
Sweden\\\\
$^2$ Laboratory of Physics, Helsinki University of Technology, FIN-02015 HUT, Finland. }

\date{\today}

\begin{abstract} 

We present explicit expressions for a large set of hierarchy wave functions on the torus. Included are the Laughlin states, 
the states in the positive Jain series, and recently observed states at \eg $\nu = 4/11$. 
The techniques we use constitute a nontrivial extension of the conformal field theory methods developed earlier 
to construct the corresponding wave functions in disc geometry. 

\end{abstract}


\maketitle

\section{Introduction}

A basic concept in the physics of the fractional quantum Hall effect (FQHE) is that of an incompressible liquid. 
Theoretically, there are two main types of such liquids depending on the quantum statistics of the quasi-particles: 
the abelian and the non-abelian liquids. Of these, only the former are firmly established experimentally. 
The abelian liquids are further divided into single- and multi-component ones, where the former 
 include the  prominent filling fractions corresponding to the  Laughlin states at $\nu = 1/(2q+1)$\cite{laughlin83} 
 and the composite fermion states in the  Jain series $\nu = n/(2qn\pm 1)$\cite{jain89}. Multicomponent liquids 
can form  in systems with spin or pseudospin degrees of freedom, \ie in partially polarized or bilayer systems \cite{multi}.

There has been two major approaches to the abelian single-component quantum Hall (QH) liquids, the 
Haldane-Halperin hierarchy \cite{haldane83,halperin84}, and the composite fermion (CF) approach of Jain 
\cite{jain89,jainbook}. In the former scheme, the quasi-particles that exist in the vicinity of a given
Laughlin state condense into a new Laughlin-like state; this procedure can be repeated thus forming an ever more
complicated hierarchy of coexisting incompressible liquids, while in the CF scheme the prominent fractions are 
understood in terms of filled effective CF Landau levels. A major advantage of the CF approach is that it provides 
explicit, and numerically very accurate, wave functions, while an appealing feature of the hierarchy scheme is 
that it treats all filling fractions on the same footing, including the newly observed ones \cite{pan} at \eg $\nu = 4/11$.

In a series of recent papers \cite{bk2,hans,CFTlong,natphys,BHHKV} we have given a concrete realization of the Haldane-Halperin 
hierarchy both by providing a large set of  explicit and testable wave functions derived using conformal field theory 
(CFT) techniques,  and by the construction of an exactly solvable model describing interacting electrons in the lowest Landau level  on a thin torus. 
The CFT construction yields candidate wave functions 
for every filling fraction obtained by successive condensation of quasielectrons (as opposed to quasiholes) 
that reduce to the exact solutions on the thin torus and coincide with those due to Laughlin and Jain, whenever these exists. 
The Laughlin states are the exact ground states of  a short range model Hamiltonian, but no such Hamiltonian is known for any other hierarchical state. Both the  Laughlin and Jain  wave functions are however known to be excellent approximations to the numerically found Coulomb groundstates for small systems. 

The standard way to quantitatively decide if  a proposed QH wave function captures the 
physics of a given phase is to study small systems numerically, typically by exact diagonalization, and 
compare the numerically obtained energies and wave functions with the proposed ones. 
To carry out such comparisons one has to choose a geometry.
Early calculations were performed in a physically motivated disc geometry, but rather  than using an explicit confining potential, 
the total angular momentum was fixed. The drawback of this method is that small systems might exhibit large, and unphysical, 
edge effects. These can be eliminated by using a finite geometry, and most large scale calculations have been performed on the sphere.
Wave functions on the torus, on the other hand, turn out to be more complicated, 
but this geometry provides other advantages. Not only are there no edges, but also no curvature, and 
since the torus is a genus one manifold, the topological 
properties of the different phases will be reflected in the ground state degeneracy \cite{wen}. 
Also, on a torus the one-dimensional nature of a Landau level is explicit: there is a natural mapping 
of the two-dimensional problem onto a one-dimensional lattice model that is solvable on the thin torus. 
There has been considerable progress in
understanding various phases of the QH system in terms of  exactly solvable models that describe the thin torus 
\cite{bk1,bk2,pfaff,seidel,read06,natphys,haldaneAPS}, for details see Ref. \onlinecite{toruslong}.

Altogether we think that there are strong reasons to construct the hierarchy  wave 
functions on the torus: it will facilitate comparisons with numerical calculations, via the ground state degeneracies 
it will give information about the topological order, and it will provide a way to  extend the exact results in the solvable 
thin torus limit to the experimentally accessible regime corresponding to a thick torus. 
Explicit wave functions on the torus have previously only been obtained in a few special cases where the structure is 
simple \cite{haldanewfs,naywil}, and in particular, the torus version of the Jain, or the hierarchy wave functions have, to our knowledge, not been constructed. 

In the present article we extend the CFT construction to obtain explicit wave functions 
in the torus geometry. In Section \ref{pre} we briefly review the necessary  CFT machinery, as well as
some technical details about the lowest Landau level on a torus. In Section \ref{wfs} we 
use this to construct wave functions on the torus. We analyze the Laughlin states (one field) and the $\nu=2/5$ state (two fields) in some detail, 
and then provide the basic relations and logical steps needed to derive the wave function for a general hierarchy state that is obtained form a Laughlin state
by successive condensations of quasielectrons. 
In Section \ref{sym} we describe an alternative choice of charge vectors that simplifies the structure of the wave functions, 
and is very useful for explicit calculations, although it is not suitable for deriving the general formulas of the previous section. 
In Section \ref{results} we  compare the $\nu=2/5$ wave function with the result of an exact numerical diagonalization 
by calculating overlaps. Section \ref{concl} includes some concluding remarks and an outlook.
Several technical steps are explained in Appendices \ref{app:bg} and \ref{app:resum}, and as a service to the practically 
oriented reader we also give explicit expressions for the states at $\nu = 3/7$ and $\nu = 4/11$ in Appendix \ref{app:expstates}.

\section{hierarchy states as CFT correlators }   \label{pre}

In this section we review the procedure, for expressing QH wave functions in a disk geometry as anti-symmetrized sums of CFT correlators, 
and emphasize several points that will be important for the generalization to the torus.
\subsection{General considerations and the Laughlin state}
Our aim is  to relate the hierarchy wave functions on the torus
to the  holomorphic conformal blocks that build the  correlation functions of vertex operators in certain CFT's.  We first briefly recapitulate the construction in the plane outlined in Refs. \onlinecite{hans, natphys}.

The relevant CFT's  have actions 
\be{action}
  S[\varphi_a]= \frac{g}{2\pi}\int \rmd^2 x\, \partial_\mu \varphi_a \partial^\mu \varphi_a,
\ee
where $\mu = 1, 2$, the metric is Euclidean, $g$ is an overall normalization to be fixed later, and 
 the boson  fields, $\varphi_a(z, \bar z)$, with $z=x_1 + ix_2 = x + iy$,  are compactified on circles with radii $R_a$.  
 In loose analogy to the standard CF construction we will 
need $n$ fields to describe the
CF state with $n$ (partially) filled CF-Landau levels
\footnote{In general, the bosons have unequal
compactification radii and hence there is in general no $O(n)$ symmetry.}.
 Note that the compactification  implies that $\varphi_a$ are angular variables, so on manifolds of higher genus, there are field configurations with nontrivial winding around the different handles.

The primary fields are given by the vertex operators,
 \begin{equation}
  \hat V_{\mbf Q }(z,\zb)= \colon e^{i \sum_{a=1}^n Q_a \varphi_a(z,\zb)}  \colon,
\end{equation}
where the colons, which we suppress  in the following, denote normal ordering, 
and $\mbf Q$ is an $n$-dimensional charge vector,
\be{chvec}
\mbf Q = \left ( \frac {e_1} {R_1}, \dots ,\frac {e_n} {R_n} \right ) ,
\ee
where $e_i$ are integers. 

On the plane we choose the normalization $g$ in the action so that the two-point function of the scalar fields is given by
\begin{equation}
  \langle \varphi_a(z,\zb)\varphi_b(w,\wb) \rangle = -\delta_{ab}\ln |z-w|^2,
\end{equation}
and hence a contraction 
of two vertex operators is given by
\begin{equation}
  \langle \hat V_{\mbf Q^{(i)} }(z,\zb) \hat V_{\mbf Q^{(j)}}(w,\wb)\rangle = |z-w|^{2  \mbf Q^{(i)}\cdot \mbf Q^{(j)} } , \label{chargevectors}
\end{equation}
for any charge vectors $\mbf Q^{(i)}$ and $\mbf Q^{(j)}$.  

The statistics of the particles is coded in the operator product expansion (OPE), or equivalently, in  the two point function,
of the corresponding holomorphic  vertex operator  $\hat V_{\mbf Q}(z)$. Setting $i = j$ in (\ref{chargevectors}), we find that the holomorphic part of 
the vertex operators with $|\mbf Q^{(i)}|^2$ odd has fermionic statistics. On the  torus, the analysis
is more involved, but the local properties of the vertex operators are the same as on the plane as is evident from the pertinent OPE's. The statistics of the 
quasiholes \cite{mooreread} and quasielectrons \cite{hans} can be understood in a similar manner.  

The QH wave functions can be obtained as correlators of the holomorphic part of the vertex operators. The simplest example is that of a Laughlin state 
at  $\nu=1/q$. Here we have only one field, $\varphi$, with radius $R^2 = q$ and charge $e=q$, and get
\begin{equation}
 \langle\prod_{i=1}^N \hat V_{\sqrt q}(z_i)
  \mathcal{O}_{bg}\rangle 
  = \Phi \prod_{i<j}(z_i-z_j)^q e^{- \frac 1 {4\ell^2} \sum_i |z_i|^2 }  \nonumber 
 \end{equation}
 \begin{equation}
   \label{eq:laughlin}
  \equiv  \Phi\    \psi_{1/q}(z_i)  ,
\end{equation}
where  $\mathcal{O}_{bg}$ is a constant neutralizing background, 
and $\Phi$ a singular phase factor that can be properly defined by replacing the continuous background charge by a lattice of thin flux tubes. 
Then $\Phi = \Phi(\phi_i(\vec n_\alpha ))$, where $\phi_i(\vec n_\alpha) $ is the  relative angle  between the coordinate $z_i$ and the lattice 
vector $\vec n_\alpha$. For the details of this procedure, which can be generalized to the full Jain series, see Appendix A  in Ref. \onlinecite{hans}.

Alternatively, we can recover the wave functions by factoring the full correlation function into a holomorphic and antiholomorphic part, also called 
conformal blocks,  each accompanied by  the square root of the non-factorizable exponential factor
\begin{equation}
 \langle\prod_{i=1}^N \hat V_{\sqrt q}(z_i,\zb_i)
  \mathcal{O}_{bg} \rangle
    = \prod_{i<j} |z_{ij}|^{2q} e^{-\frac 1 {2\ell^2}\sum_i |z_i|^2 }  
  \nonumber\end{equation}
 \begin{equation}
 = \left(  \prod_{i<j}z_{ij}^q e^{ -\frac 1 {4\ell^2} \sum_i |z_i|^2 } \right)^\star \left(
  \prod_{i<j}z_{ij}^q e^{- \frac 1 {4\ell^2} \sum_i |z_i|^2 } \right)
\nonumber  \end{equation}
  \begin{equation}
    \label{eq:laughlin2}
\equiv (\psi_{1/q}(z_i))^\star  \psi_{1/q}(z_i)         \,  ,
\end{equation}
where $z_{ij}=z_i-z_j$. 
It is this latter procedure that will generalize to the torus. The background charge must be included to obtain a non-vanishing correlator, 
and will be treated in detail in Section \ref{wfs}. For now it suffices to mention that it gives rise to the correct nonholomorphic dependence, 
and nothing more, as it does in the disc geometry. 

On the torus, correlation functions will not factorize as in \pref{eq:laughlin2}, but are given by an (in general infinite) sum over such terms. 
For a special class of CFT's, this infinite sum can be rewritten as a finite sum over extended conformal blocks. Such theories are called 
rational CFT's \cite{yellowbook}. For a bosonic action of the type \pref{action} this requires that the radii, $R_a$, are of the form 
$R_a^2 = 2p/p'$ with $p$ and $p'$ relatively prime. This 
particular rational CFT is called a rational torus.  Clearly the Laughlin state considered above falls into this class, as do the hierarchy states 
that are discussed in the following section.  

In the general case we will thus extract not only a single wave function, but a whole set. This provides an important consistency check 
on our method, since the  degeneracy of a lowest Landau level (LLL) state on the torus is known 
from the general symmetry analysis due to Haldane  \cite{haldanetorus}. 
The degeneracy of a state with filling fraction $\nu = p/q$,  includes a factor four related to choosing  periodic or antiperiodic 
boundary conditions along the cycles of the torus, and a $q$-fold degeneracy related to the position of the center-of-mass.  
As we will see, our construction precisely recovers this degeneracy.

\subsection{The hierarchy wave functions}\label{hwf}

The hierarchy wave functions at level $n$,  obtained by successive  condensation of quasiparticles only,  are described  by a set of integers $\{k_1,\ldots,k_n\}$, 
related to the densities of the quasiparticle condensates. 
 In the CFT scheme of Ref. \onlinecite{natphys} these wave functions are constructed from correlators involving $n$ different vertex operators, $V_\alpha$. 
They depend on  the boson fields $\varphi_a \, , \ a=1,\dots, \alpha$ and are constructed recursively by
\be{start}
V_1&=& e^{i{\gamma_{1} \varphi_{1}}}     \nonumber \\
V_{\alpha+1}&=&\partial_z V_{\alpha} e^{-i{ \varphi_{\alpha}/ \gamma_\alpha}} e^{i{\gamma_{\alpha +1} \varphi_{\alpha +1}}}  ; \ \alpha=1,\dots, n-1
\ee
where  $\gamma_1=\sqrt{2k_1+1}$, $\gamma_{\alpha+1}= \sqrt{2k_{\alpha+1}-\gamma_ \alpha ^{-2}}$ and $\varphi_{\alpha +1}$ is a new bosonic field. 
The electrons are divided into $n$ sets $I_\alpha$ of size $M_\alpha$ and the  wave function is given by  
\be{wfexpr}
\Psi ={\cal  A}  \langle\prod _{\alpha=1} ^{n}\prod_{i_\alpha \in I_\alpha} 
V_\alpha (z_{i_\alpha}) \rangle \, ,
\ee
 where $\cal A$ denotes antisymmetrisation and $\av{\dots}$  the correlation function in a suitable background field. The  $M_{\alpha}$ are 
 determined recursively as explained in Ref. \onlinecite{BHHKV}.
The filling factor of $\Psi$, $\nu_n=p_n/q_n$, is obtained recursively from
\be{fillfr}
\nu_n=\frac{p_n}{q_n}=\frac{2k_np_{n-1}-p_{n-2}}{2k_nq_{n-1}-q_{n-2}} \, , 
\ee
 with the  initial conditions $q_0=p_0=1$,  and the Laughlin series given by $\nu_1={1} /({2k_1+1})$, $k_1$ integer.  
Taking $k_j=1$ amounts to having maximal density in the $j^{\rm{th}}$ condensate, and the Jain series $\nu_n={n}/({2nk_1+1})$ corresponds to  choosing $k_2=\dots =k_n=1$.

To evaluate the correlators of the operators \pref{start} on the torus, we will use a different representation of the vertex operators, where the charge vectors are explicit:   
 \be{vertop}
 V_{ \mbf Q^{(\alpha ,n)} } (z,\bar z) =\mathcal{D}^{(\alpha-1)} \hat  V_{ \mbf Q^{(\alpha ,n)} } \, .
 \ee
 Here, $\mathcal{D}^{(\alpha)}$ are derivative operators to be discussed below. 
A vertex operator without a hat is a descendant, whereas a hat marks a primary field.  
 The charge vectors  $\mbf Q^{(\alpha,n)}$ are not uniquely defined. We will mostly use a parametrization where the hierarchy construction is manifest:
\be{chargevector}
\mbf Q^{(\alpha,n)}=(\frac{c_1}{R_1},..., \frac{c_{\alpha-1}}{R_{\alpha-1}},\frac{q_{\alpha}}{R_\alpha},0,...,0).\ee 
The CFT charges, $c_j$ and $q_j$, and the compactification radii, $R_j$,  are connected to the denominators of the filling fractions by 
\be{cqr}
c_j=q_j-q_{j-1} \ , \ \ \  R_j^2=q_j q_{j-1} \ . 
\ee
For the Jain series, for example,  all charges are equal to $2k_1$.  When there is no ambiguity, we suppress the level index $n$ on the charge vector and write just $ \mbf Q^{(\alpha)} $.

Other choices of basis sets represent the FQH state equally well as long as the bosonic fields, $\varphi_1,\dots,\varphi_n$, have rational $R_i^2$ 
and the inner product of two vectors is unchanged: $\mbf Q^{(\alpha)} \cdot \mbf Q^{(\alpha')}= \delta_{\alpha\alpha'}+2k_1+\sum_{j=2}^\alpha 2(k_j-1)$, 
for $\alpha\leq \alpha'$. In the special case of the Jain series, this relation simplifies to: $ \mbf Q^{(\alpha)} \cdot\mbf Q^{(\alpha')}=\delta_{\alpha \alpha'}+2k_1$ \cite{hans}. 
This is discussed in more detail in Section  \ref{sym}. 

For later computations in Section \ref{sechierarchy}, we will also need the following relation between the number of states, $N_s$, and the charge vectors:
\be{Ns-rel}
\forall \alpha\ : \ \sum_{\beta=1}^n \mbf Q^{(\alpha)}\cdot\mbf Q^{(\beta)} M_\beta=N_s,
\ee
 Together with $\sum_\beta M_\beta=N=\nu N_s$,  this  gives a consistency condition on $N_s$.

\section{QH states on the torus - general considerations}  \label{wfs}
In this section we describe the techniques we use  to construct the hierarchy wave functions  on the torus. 

\subsection{Lowest Landau level on the torus}
We consider a system of $N$ charges $-e$ on the torus with periods $L_1$ and $L_2$.
In Landau gauge, a homogeneous external magnetic field perpendicular to the surface is described in terms of a vector potential $\vect{A} = By \hat{\vect{x}}$.
Most of the time we will 
set the magnetic length equal to unity, $\ell=\sqrt{\hbar c/eB} = 1$

In the presence of an external magnetic field, consistent periodic boundary conditions can only be enforced up to a gauge transformation.
A natural way to do this is to use the magnetic translation operators \cite{haldanetorus} 
\begin{equation}
  t(\mbf l) =e^{ \mbf l\cdot( \nabla - i \mbf A)- i  \mbf l \times \mbf x} \, ,
\end{equation}
where $\vect{l}$ parametrizes the translation, and $\vect x$ the electron coordinate. 
In Landau gauge we define
\begin{align}
  &t_1 \equiv t(\tfrac{L_1}{N_s} \hat{\mbf x})  = e^{\frac{L_1}{N_s}\partial_x},
  &t_2 \equiv t(\tfrac{L_2}{N_s} \hat{\mbf y})  = e^{\frac{L_2}{N_s}(\partial_y+i x)},
\end{align}
where $N_s=L_1L_2 /2\pi \in \mathbb{Z}$ is the number of flux quanta penetrating the surface of the torus.
The translations $t_1^{N_s}$ and 
$t_2^{N_s}$ commute. Hence, periodic 
boundary conditions can be consistently formulated as
\begin{equation}
  \label{eq:pbc}
  t_\mu^{N_s}\psi = \exp\left(i \phi_\mu \right)\psi     \ \  ; \ \      \mu=1,2\,\,,
\end{equation}
where $\phi_\mu$ can be interpreted as solenoid fluxes through the handles of the torus.
Once the fluxes are fixed, the same boundary conditions apply for all states
in the Hilbert space of the system. 

A complete set of  commuting operators for a $\nu=p/q$ QH system, 
with $p$ and $q$ relative primes, is given by $\{H,T_1,T_2^q\}$,
where $T_\mu=\prod_{i=1}^N t_{\mu,i}$ are center-of-mass translations and $H$ is a translationally invariant electron-electron interaction. 
The states are labeled by $\{E, K_1,K_2\}$ where 
the quantum numbers $K_\alpha$ of the system relate to the eigenvalues of $T_1$ and $T_2^q$, which are given by $e^{2 \pi i K_\alpha/N_s}$  respectively. 

The lowest Landau level wave functions, in Landau gauge, on the torus
have many special properties discussed for instance in 
Refs. \onlinecite{haldanetorus,read96}. They all factorize as
\begin{equation}
  \psi(z_1,\ldots,z_N) = f(z_1,\ldots,z_N)e^{-\sum_{k=1}^N y_k^2/2},
\end{equation}
where $f$ is holomorphic in all its arguments and is given in terms of generalized $\vartheta$-functions
\begin{equation}\label{Jacobi}
  \vartheta\begin{bmatrix} a\\b\end{bmatrix}(z|\tau)
  = \sum_{k=-\infty}^\infty  e^{i\pi\tau(k+a)^2}e^{2\pi i(k+a)(z+b)}.
\end{equation}
For example, a Jastrow-type factor $\prod_{i<j} (z_i-z_j)^q$ on the plane becomes $\prod_{i<j}\vartheta_1(\frac{z_i-z_j}{L_1}|i\frac{L_2}{L_1})^q$ on the torus, where $\vartheta_1$ corresponds to $a=b=1/2$.

From the boundary conditions in Eq. \ref{eq:pbc}, one derives the corresponding
conditions on the holomorphic functions $f$ as will be discussed in detail below. 

Because of translational invariance, the 
center-of-mass (CM)  dependence of the wave function can be separated:
\begin{equation}
  f(z_1,\ldots, z_N) = F_{cm}(Z)f_{rel}(z_1,\ldots, z_N),
\end{equation}
where $Z=\sum_i z_i/L_1$ and $f_{rel}$ is independent of $Z$.
While for the Laughlin states $f_{rel}$ is uniquely determined, 
(at least on the plane, sphere and torus), by its leading short-distance behaviour, this is not
the case for the Jain states, or more general hierarchy states.
Of course, once the relative part is fixed, the degeneracy of the corresponding level is given by the number of center-of-mass functions compatible with the boundary conditions.

For hierarchy wave functions, we have not been able to explicitly separate the CM part even in the simple case of the Jain sequence. 
Here, the CFT techniques have proved to be extremely useful in that they allow for a direct construction of the wave functions without ever separating the CM part.

\subsection{The background charge} \label{bgcharge}
A careful calculation shows that the correlators
$\langle\prod_i V_{\mbf Q^{(i)}}(z_i,\zb_i) \rangle$
vanish unless they satisfy the 
charge-neutrality condition $\sum_i \mbf Q^{(i)} =\mbf  0$. 
On the plane, there are two standard choices for the neutralizing background.
For $N$ particles with charge vectors, $\mbf Q_a$, one can either assume a compensating charge at infinity, 
$ \mathcal{O}_{bg} = e^{-i N \sum_{a}  Q_a \varphi_a(z_\infty,\zb_\infty)} $
or a homogeneous droplet, 
\be{bgop}
\mathcal{O}_{bg} = \prod_{a=1}^n     \exp[-i R_a \rho_a\int \rmd^2x\,\varphi_a(z,\bar z)] \, ,
\ee
with $\rho_a={\rho_0}/{R_a^2}$, $\rho_0={1}/({2\pi \ell^2})$, being the density of the $a^{\rm th}$ set. 
The latter method, which is more physical, is the one used in \pref{eq:laughlin2}; it reproduces the correct non-holomorphic dependence of the LLL wave functions in the symmetric gauge. 
On the torus, the first choice is not possible at all, therefore we will use the latter. 

As in the plane, this homogeneous distribution cannot be realized using the 
operator content of the bosonic rational CFT, which implies that, in general,
we cannot expect the conformal correlation functions to factorize into 
chiral components. On the plane, this did not cause any problems. On the contrary,  the homogeneous background charge actually 
contributed the gaussian non-holomorphic dependence needed in the LLL wave functions. Below, we find the same result on the torus.

\subsection{The derivative operators $ \mathcal{D}^{(k)}$  }

One of the basic difficulties in translating the hierarchy wave functions from the plane to the torus is related to the meaning of the 
derivative operators $ \mathcal{D}^{(k)}$. In the plane, they are the holomorphic partial derivatives $\partial_z^k$ implying that 
the operators  $V_{\alpha}$ are descendants of the primary fields in the theory. Using standard techniques, the corresponding 
correlation functions can be written as a product of derivatives acting on the correlation function of primary fields only.  As discussed 
in  Ref. \onlinecite{hans}, the derivatives act only on the polynomial part of the wave function, that is on the conformal block, and not 
on the exponential factor. Thus, in the plane, the calculation of the wave functions reduces to the calculation of correlators of primary fields.

The origin of the derivative operators is easily understood in terms of composite fermions, where they are the remnants of the effective Landau level structure. Projection onto the lowest Landau level converts the $\bar z_i$'s, which occur in higher Landau levels,  into derivatives, $\partial_i$. In particular, this means that if a derivative is excluded, the wave function will vanish since it amounts  to putting a particle in an already filled Landau level. For the general hierarchy state, the situation is more complicated,  but  
numerical calculations indicate that the derivatives are necessary also here. 

Turning to the torus, we have not been able to prove that the derivatives are needed in order to get non-vanishing wave functions. On the other hand,  this is almost obvious since for large tori, the wave functions must be very similar to those on the plane, and these do need derivatives. We have also checked that the correlators of the primary fields do vanish after antisymmetrization in simple cases, \eg for $\nu = 2/5$.

A natural way to construct our wave functions on the torus, would thus be to supplement an appropriately  periodized version of the Jastrow factors occurring in the disc version, with derivatives and a suitably chosen CM part. Such an ansatz, however, does not work.  A derivative acting only on the holomorphic part of a correlator destroys the quasi-periodicity of the wave function. \footnote{
An obvious way to preserve  the correct symmetry would be to introduce a suitably choosen
covariant derivative. This will however  necessarily introduce $\bar z_i$'s in the wave functions, thus reintroducing the need for projection and destroying an appealing feature of our approach.
More importantly, in the simplest cases like $\nu =2/5$, it turns out that the wave functions are entirely in the second LL, and thus vanish after projection to the LLL 
on the torus.
}

We conclude that we must construct the $\mathcal D^{(k)}$'s in \pref{vertop} from operators that preserve the Landau level index and the boundary conditions, {\it ie} they must commute with the one-particle hamiltonian and the magnetic translations, $t_{\mu ,i}^{N_s}$, around the cycles of the torus.  Requiring, in addition, that they  approach 
$a\partial + b\bar \partial + c$ (where $a,b,c$ are constants)  on  a large torus, leads to the  finite translations in the $x$-direction:
\be{derop}
%\mathcal{D}^k_i=t_{1i}^k = e^{\frac {k L_1} {N_s} \partial _{x_i}} \, .
t_{1i}^k = e^{\frac {k L_1} {N_s} \partial _{x_i}}    =  e^{\frac {2\pi k \ell^2} { L_2}  ( \partial_i + \bar\partial_i ) }   \, ,
\ee 
where there are at most $N_s$ distinct possibilities for the integer $k$. We note that these operators will preserve the quantum numbers of the many-body wave functions.
A more thorough analysis shows that, in fact, only $\lceil{N_s}/{2}-1\rceil $---that is, ${N_s}/{2}-1$ if $N_2$ is an even number and $(N_s-1)/{2}$ if $N_s$ is odd---are independent.
Forming  $\mathcal D^{(k)}$ from linear combinations of  the operators \pref{derop},
\be{lincom}
 \mathcal{D}^{(k)} = \sum_{m=0}^{\lceil{N_s}/{2}-1\rceil } c^{(k)}_m t_1^m \, ,
\ee 
we can first compute the correlators of the primary fields, and then act with the $\mathcal D^{(k)}$'s, just as in the plane. 
To get the correct limit for a large torus, it is natural to demand that $\mathcal D^{(k)} \rightarrow \partial^k$ as $N_s= L_1L_2/(2\pi \ell^2) \rightarrow \infty$, but this is not enough to determine the coefficients $c^{(k)}_m$  uniquely, and may in fact not even be necessary.\footnote
{
We have in fact found that taking $D^{(1)} =  t_1, D^{(2)} =  t_1^2$ gives a wave function identical to the one obtained using \pref{deran}. 
We do not have an analytical understanding of this, but it does suggest the very simple ansatz $\mathcal{D}^{(k)} =  t_1^k$.
}
For the states at level two and three that we have tested, we used,
\be{deran}
\mathcal D^{(1)} &=&  t_1 - 1 \\
\mathcal D^{(2)} &=&  t_1^2 - 2t_1 + 1 \, .  \nonumber 
\ee
(The constant in $\mathcal{D}^{(k)}$ can be ignored as it gives no contribution to wave functions.)
That the $c^{(k)}_m$'s cannot be determined uniquely might seem a  serious drawback, but in Section \ref{results} where we present numerical tests of our wave funtions, we will see that for the $\nu=2/5$ state, the simplest choice, $\mathcal D^{(1)}=  t_1$, already gives a good result. Furthermore, we show how this can be improved to yield a very good wave function  by taking a linear combination 
of states where $ \mathcal{D}^{(1)} = t_1^k $, $k$ small.
We believe that adding such contributions from higher values of $k$ is the torus counterpart of correcting the wave functions in the plane by adding contributions involving descendants of the primary fields defining the representative wave functions \cite{mooreread}.

\subsection{Expressions for the correlators}
Following the strategy outlined above we now ignore the derivative operator in $V_\alpha$, and calculate the correlators of the corresponding primary operators, $\hat V_\alpha$, in the presence of a homogeneous background. 
Since the scalars $\varphi_a$ are not coupled,  the correlation function  \pref{wfexpr}  factorizes and can be calculated using  a straightforward
generalization of known techniques (see \eg Chapter 12 of Ref. \onlinecite{yellowbook}):
\be{factor}
 & \langle& \prod _{\alpha =1} ^{n}\prod_{i_\alpha \in I_{\alpha} }
\hat V_{\mbf Q^{(\alpha)} } ( z_{i_\alpha} ,  \bar z_{i_\alpha}  ) \mathcal{O}_{bg} \,  \rangle  \nonumber\\
&=&    \langle \ \prod _{\alpha =1} ^{n}\prod_{i_\alpha \in I_{\alpha} }
e^{i\sum_a Q_a^{(\alpha)} \varphi_a  ( z_{i_\alpha} ,  \bar z_{i_\alpha}  ) }\mathcal{O}_{bg} \  \rangle \nonumber \\
&=& \prod_{a=1}^n    \langle  \ \prod _{\alpha =1} ^{n}\prod_{i_\alpha \in I_{\alpha} }
e^{i Q_a^{(\alpha)} \varphi_a  ( z_{i_\alpha} ,  \bar z_{i_\alpha}  )} \mathcal{O}_{bg} \ \rangle \nonumber \\
&=& |\psi_{jas} (z_{ij})|^2   \prod_{a=1}^n  {\cal F}^{(a)} (Z^{(a)} ,\bar Z^{(a)} )   f_{bg}^{(a)} (z_i, \bar z_i) 
\ee
with
 \be{part1}
  \psi_{jas} (z_{ij}) &=& 
  \prod_{\gamma=1}^n\prod_{i_\gamma < j_\gamma \in I_\gamma} 
\vartheta_1( z_{i_\gamma j_\gamma} |\tau)^{ \mbf Q^{(\gamma)} \cdot \mbf Q^{(\gamma)}} \nonumber \\
 &\times&
 \prod_{\alpha < \beta} ^n \prod_{i_\alpha\in I_\alpha \atop i_\beta \in I_\beta }
 \vartheta_1(z_{i_\alpha i_\beta}  |\tau)^{ \mbf Q^{(\alpha)}\cdot  \mbf Q^{(\beta)}  }  \, ,
\ee
\be{part2} 
{\cal F}^{(a)} (Z^{(a)} ,\bar Z^{(a)} )  =  \!\!\!\!\sum_{e,m=-\infty}^\infty \!  \!\! e^{-2\pi i \rho_a R_a \int  \frac {\rmd^2x}{L_1}   \,\left[\alpha_{e,m}z-\bar\alpha_{e,m}\bar z \right] }  \nonumber \\
 \times  e^{i\pi \left[ \tau \alpha_{e,m}^2  -\bar\tau\bar\alpha_{e,m}^2 \right] }
  e^{2\pi i\left[ \alpha_{e,m} \frac { Z^{(a)} } { R_a  } -  \bar\alpha_{e,m}\frac { \bar Z^{(a)} } {  R_a } \right]} 
  \ee
  and
  \be{part3}
  f_{bg}^{(a)} (z_i, \bar z_i) \hskip 6cm   \\
= \prod_{\alpha=1}^n \prod_{i_\alpha\in I_\alpha} 
  e^{-\rho_a R_a  Q^{(\alpha)}_a\int\rmd^2 x\, \ln |\vartheta_1(  \frac {z-z_{i_\alpha}} {L_1} |\tau) |^2 } \nonumber
 \, ,
\ee
where  $z_{i_\alpha j_\beta} = (z_{i_\alpha} - z_{j_\beta})/L_1$ {\it etc.}, and 
\be{not}
Z^{(a)}&=&R_a \sum_{\alpha=1}^n Q_a^{(\alpha)}{Z_\alpha}\nonumber\\
&=&q_aZ_a+c_a\sum_{\alpha=a+1}^n Z_\alpha \, ,
\ee
with $Z_\alpha = \sum_{i_\alpha \in I_\alpha} z_{i_\alpha} /L_1$ being the (dimensionless) CM coordinate of the electrons in  the set $I_\alpha$. We also introduced the notation 
 $\tau=i L_2/L_1$  for the modular parameter describing  the torus, and  
$\alpha_{e,m}=e/R_a+m R_a/2$, $\bar\alpha_{e,m}=e/R_a-m R_a/2$, where $e,m$ are integers, to parametrize the 
electric and magnetic sectors of the CFT Hilbert space. 
Note that $Z^{(a)}$ depends on the filling fraction $\nu_n$. In order to avoid confusion, we will  often indicate the level of hierarchy by the subscript $n$, $Z_n^{(a)}\equiv Z^{(a)}$.  When an explicit expression is  needed we will instead denote it by the filling fraction, like \eg $Z^{(a)}_{2/5}$.

The first exponential in \pref{part2} comes from the effect
of the background charge for the different fields and vanishes, if the integration domain is chosen to be
$\int d^2 x\, \equiv \int_{-L_1/2}^{L_1/2}dx \, \int_{-L_2/2}^{L_2/2}dy$. 
The integral in \pref{part3} can be calculated exactly, and using the same integration domain, we obtain,
\be{bgcon}
  I(z,\bar z) = \int \rmd^2x' \ln \left| \vartheta_1(\frac{z'-z}{L_1}|\tau) \right|^2
  = I(0,0) + 2\pi y^2.
\ee
In appendix A, we show that a different choice 
of integration domain only amounts to a coordinate shift in the wave function. 
This, however, is not obvious, but  a rather non-trivial consequence of the homogeneity of the states. 

Substituting \pref{bgcon} in \pref{part3} yields, up to a constant factor,
\be{bgcon2}
  f_{bg}^{(a)} (z_i,\bar z_i)  =  \prod_{\alpha=1}^n \prod_{i_\alpha\in I_\alpha} e^{ -2 \pi R_a \rho_a  Q^{(\alpha)}_a y_{i_\alpha}^2} \, .
 \ee
Finally we can use the relation $ \sum_{a=1}^n   Q^{(\alpha)}_a/R_a =1$, which simply
expresses that all electrons have unit charge \cite{hans}, to get
 \be{gauss}
\prod_a f_{bg}^{(a)} = e^{-\sum_{k=1}^N y_k^2/\ell^2} \, ,
 \ee
 which is exactly the non-holomorphic  gaussian factor appropriate to Landau gauge.

Since the background charge does not alter the form of the 
charge-lattice summation the conformal blocks have the same structure as 
in a rational CFT.
We stress that because of the gaussian factor $f_{bg}$, and that alone, the terms in the sum that give the full correlator,  cannot be factorized into holomorphic and antiholomorphic blocks as it would 
should the operator content be purely that of a  rational CFT. As in \pref{eq:laughlin2}, we extract the QH wave function as the holomorphic conformal block times the square root of the gaussian factor.
In the following, we assume a homogeneous  neutralizing background and only write the fully (anti)holomorphic blocks omitting the gaussian factor.

\subsection{Charge lattice sums }
In order to diagonalize magnetic translations in the Hilbert space spanned
by the holomorphic conformal blocks, it is useful to simplify the charge-lattice
sum. If the compactification radius is of the form $R_a^2=2p/p'$, then  we can express the infinite sum in \eqref{part2} as a finite sum over extended conformal blocks (see Appendix \ref{app:resum}) 
\begin{multline}
  \label{eq:resummed}
  \sum_{e,m} e^{i\pi\tau \alpha_{e,m}^2} e^{-i\pi\bar\tau\bar\alpha_{e,m}^2}
  e^{2\pi i\left[ \alpha_{e,m}Z^{(a)}-\bar\alpha_{e,m}\bar Z^{(a)}\right]/R_a}
  \\= \sum_{r=0}^{p'-1}\sum_{s=0}^{2p-1}\mathcal{F}_{r,s}(Z^{(a)})
  \bar{\mathcal{F}}_{-r,s}(\bar Z^{(a)})\, ,
\end{multline}
where the $2pp'$ functions
\be{eq:Fblock}
  \mathcal{F}_{r,s}(Z^{(a)}) &=& \sum_{k=-\infty}^\infty e^{ i \pi \tau (2pp'k+r p +s p')^2/2pp'}  \nonumber \\
  &\times& 
  e^{ \frac {i\pi} {p} (2pp'k+r p+s p') Z^{(a)}  } 
\ee
span the Hilbert space of center-of-mass motion. Under single-particle lattice
translations they transform as follows:
\begin{align}
  \label{eq:F_transf}
 \frac{ \mathcal{F}_{r,s}(Z^{(a)}+c)}{\mathcal{F}_{r,s}(Z^{(a)})}&= (-1)^{c r}e^{2\pi i  c s/R_a^2}\nonumber\\
  \frac{ \mathcal{F}_{r,s}(Z^{(a)}+c\tau)}{ \mathcal{F}_{r,s+c}(Z^{(a)})} &=e^{-i\pi\tau c^2/R_a^2}e^{-2\pi i cZ^{(a)}/R_a^2}\, .
\end{align}
Here  $c$ are the integers related to the charges in the charge vectors \eqref{not}.

\section{QH states on the torus -  explicit constructions}  \label{wfc2}

We now have all the pieces needed to extract the wave functions---except for one essential ingredient: if we extract the basis  functions $\psi_s$ from the correlators using the recipe  $\sum_s [\psi_s ( z_i)]^\star \psi_s(z_i)$ discussed in Section \ref{pre}, these will not satisfy the boundary conditions \pref{eq:pbc}. 
In this section, we describe in detail how to construct the correct linear combinations for the hierarchy states at filling fractions $\nu_n$  \eqref{fillfr}.

We start by constructing the wave functions explicitly in two simple cases: First the Laughlin state at $\nu = 1/(2k +1)$ and then the simplest level two state, the $\nu = 2/5$ in the Jain sequence.  The reader who carefully studies these examples should get a pretty good idea about the strategy for attacking the general case, the details of which are given in the last subsection. 

\subsection{The Laughlin state}\label{laughlin}
As already discussed, the Laughlin wave function for  $\nu=1/q$ on the torus \cite{haldanewfs} can be extracted from 
the $N$-point correlation function of the vertex operators 
$V_{\mbf Q^{(1)}}$ depending on a single field $\varphi_1$ compactified on a circle with radius  $R_1^2=q$ (\ie $p=q$ and $p'=2$), and with charge vector $ \mbf Q^{(1)} = q/R_1 = \sqrt{q}$. In this case there is only one condensate and $Z^{(1)} =  3 Z$. 

Using \eqref{factor} and \eqref{eq:resummed}
we can express the holomorphic wave functions in a basis
given by the chiral conformal blocks,
\begin{equation}
  \label{eq:laughlinbasis}
  \psi_{r,s}(z_i)= \prod_{i<j}\vartheta_1(z_{ij}|\tau)^{q}
  e^{-\tfrac{1}{2}\sum_k y_k^2}    \mathcal{F}_{r,s}(Z^{(1)})\, .
\end{equation}
A maximal linearly independent set of states is obtained with 
$r=0,1$ and $s=0,1,\ldots, 2q-1$. 
The number of conformal blocks is $2pp' = 4q$ which is the expected number of degenerate states---a factor $q$ coming from the CM degeneracy, and a factor $2\times 2$ from the different boundary conditions \pref{eq:pbc}.
The $q$-fold degenerate multiplet of
Laughlin wave functions is given by the linear combinations of the basis states, \eqref{eq:laughlinbasis}, that transform irreducibly under magnetic lattice translations  and satisfy the same solenoid flux conditions. 
 
To find the physical states we have to diagonalize the action 
of the single particle magnetic translation operators, $t_{\alpha,i}^{N_s}$, on the full wave functions. To make the analysis more transparent,
however, let us first summarize the relevant transformation properties of 
the holomorphic components of the basis states. The odd Jacobi
theta function transforms as
\be{jactheta}
  \vartheta_1(z+1|\tau) &=&-\vartheta_1(z|\tau)\nonumber\\
  \vartheta_1(z+\tau|\tau)&=& -e^{-i\pi\tau}e^{-2\pi i z}
  \vartheta_1(z|\tau) \, .
\ee
The transformation properties of the holomorphic center-of-mass functions
can be read off from \eqref{eq:F_transf} and are given by
\begin{align}
  \mathcal F_{r,s}(Z^{(1)}+q) &=(-1)^r \mathcal F_{r,s}(Z^{(1)})\nonumber \\
  \label{eq:Ftau}
  \mathcal F_{r,s}(Z^{(1)}+q \tau)  &=e^{-i\pi\tau q}e^{-2\pi i Z^{(1)}}\mathcal F_{r,s+q}(Z^{(1)}) \, ;
\end{align}
the functions should also satisfy the
periodicity conditions
\begin{align}
  \mathcal F_{r,s+2q}(Z^{(1)}) &= \mathcal F_{r,s}(Z^{(1)}),\nonumber\\
  \mathcal F_{r+4,s}(Z^{(1)}) &= \mathcal F_{r,s}(Z^{(1)}).
\end{align}

From these relations, we can work out the action of the magnetic translations 
on the basis functions \eqref{eq:laughlinbasis};  in the $x$-direction this gives:
\begin{align}
  \label{eq:act_t1}
  \frac{t_{1,i}^{N_s}\psi_{r,s}(z_i)}{\psi_{r,s}(z_i)} &= 
  \prod_{i < j}\frac{\vartheta_1(z_{ij}+1|\tau)^{q}}{\vartheta_1(z_{ij}|\tau)^{q}}
  \frac{\mathcal F_{r,s}(Z^{(1)}+q)}{\mathcal F_{r,s}(Z^{(1)})}\nonumber\\ &= (-1)^{q(N-1)+r} \, ,
\end{align}
and in the $y$-direction:
\begin{multline}
  \frac{t_{2,i}^{N_s} \psi_{r, s}(z_i) } {\psi_{r, s+q} (z_i)}=\\ 
\prod_{j\neq i}  \frac{\vartheta_1(z_{ij}+\tau|\tau)^{q}}{
    \vartheta_1(z_{ij}|\tau)^{q}}
  \frac{\mathcal F_{r, s}(Z^{(1)}+q \tau)}{\mathcal F_{r, s+q}(Z^{(1)})}
  e^{i \,\imag \tau z_i +\tfrac{1}{2}(\imag \tau)^2}\\
  =(-1)^{q(N-1)} ,
\end{multline}
where, in the last step, we used the identities 
$Nq= N_s$ and $ L_1L_2= 2\pi N_s$. 
This shows that $\psi_{r,s}$ are $t_{1,i}^{N_s}$ eigenstates with 
eigenvalues independent of $s$, while $t_{2,i}^{N_s}$ maps the 
function $\psi_{r, s}(z_i) $ into the
linearly independent function $\psi_{r, s+q}(z_i) $. 
Since, however, both transformations are independent of $s$, we can, in a unique way, 
satisfy  the boundary conditions \pref{eq:pbc}  by forming the linear combinations 
\be{Firr}
  {\cal H}_{r,t,\bar s}^{(1)} (Z^{(1)})&=& \mathcal F_{r,\bar s}(Z^{(1)})+(-1)^t \mathcal F_{r,\bar s+q}(Z^{(1)}) \\
  &=&
 \sum_k (-1)^{kt}e^{ i\pi \tau q(k+a_1)^2}e^{2\pi i(k+a_1)q Z} \nonumber \, ,
 \ee
where $a_1=\frac{\bar s}{q}+\frac{r}{2}$, $t = 0,1$ and $\bar s = 0,\ldots, q-1$, which amounts to a change of basis
for the conformal blocks spanning the Hilbert space of the CM wave function. 
Thus, defining
\be{findef}
\psi_{r,t,\bar s} (z_i) = 
{\prod_{i < j }  \vartheta_1(z_{ij}|\tau)^{q} }  {\cal H}_{r,t,\bar s}^{(1)}   (Z^{(1)})   
  e^{-\tfrac{1}{2}\sum_k y_k^2}  
\ee
we finally have
\be{finbc}
t_{1,i}^{N_s} \psi_{r,t,\bar s} (z_i)
 &=& (-1)^{q(N-1)+r}\psi_{r,t,\bar s} (z_i)  \nonumber\\
t_{2,i}^{N_s}\psi_{r,t,\bar s} (z_i)   
  &=&(-1)^{q(N-1)+t} \psi_{r,t,\bar s} (z_i) \, . 
\ee
Thus, the states $\psi_{r,t, \bar s}$ are eigenstates of $t_{\alpha,i}^{N_s}$, $\alpha=1,2$.
For a fixed number of particles, the four different choices of the  solenoid fluxes in Eq. \ref{eq:pbc} 
precisely correspond to the four combinations of the quantum numbers $r$ and $t$, while the
quantum number $\bar s$ determines the position of the CM.
For a translationally invariant hamiltonian, this implies a $q$-fold degeneracy.

The many-body  quantum numbers
are related to the CM translations 
$T_1=\prod_{i=1}^N t_{1,i}$ 
and $T_2^q=\prod_{i=1}^N t_{2,i}^q$ which, together with the 
hamiltonian, form a maximal set of commuting operators. 
The operator $T_1$ acts only on the holomorphic CM piece, and we get
\begin{eqnarray}
  T_1 \mathcal{H}^{(1)}_{r,t,\bar s}(Z^{(1)}) &=& \mathcal{H}^{(1)}_{r,t,\bar s}(Z^{(1)}+L_1) \nonumber \\ %= F_{r,s}(Z+\frac{1}{q})
  &=& (-1)^r e^{2\pi i s/q}\mathcal{H}^{(1)}_{r,t,\bar s}(Z^{(1)}).
\end{eqnarray}
Comparing with the definition of the quantum numbers $K_i$, given by $T_i\psi = e^{2\pi i K_i/N_s}\psi$,  it follows that $K_1 = (rN_s/2+N \bar s )\mod N_s$. 
$T_2$ acts on both the gaussian and on the holomorphic
CM piece. We get
\begin{multline}
  T_2^q\left[e^{-\tfrac{1}{2}\sum_k y_k^2} \mathcal{H}^{(1)}_{r,t,\bar s}(Z^{(1)})\right]
    \\ =
    (-1)^te^{-\tfrac{1}{2}\sum_k y_k^2} \mathcal{H}^{(1)}_{r,t,\bar s}(Z^{(1)}),
\end{multline}
and hence $K_2= (tN_s/2) \mod N_s$.  
This analysis establishes that 
states with unequal $\bar s$ have different quantum numbers and are therefore orthogonal.
We conclude, in agreement with Ref. \onlinecite{haldanewfs}, that
the degeneracy of Laughlin's states, as obtained from conformal correlators,
is given by the denominator of the filling factor.

Let us summarize the result for the case $\phi_a=0$, where $r=t=(N_s-q)\mod 2$. In order to write the result
in a more conventional notation, we note that
the holomorphic CM function can be written in terms of
the Jacobi theta function \eqref{Jacobi}.
Comparison with \eqref{Firr} gives
\begin{equation}
\mathcal{H}^{(1)}_{r,t,\bar s}(Z^{(1)}) = \vartheta
    \begin{bmatrix}
      (N_s-q)/2+\bar s/q\\
      (N_s-q)/2
    \end{bmatrix}(qZ|q\tau).
\end{equation}
In this notation, the $q$-fold degenerate multiplet of Laughlin wave 
functions is given by
\begin{multline}
  \psi_{\bar s}(z_i) =  \prod_{i<j}^N\vartheta_1(z_{ij}|\tau)^q
  e^{-\frac{1}{2}\sum_k y_k^2} \\ \times \vartheta
    \begin{bmatrix}
      (N_s-q)/2+\bar s/q\\
      (N_s-q)/2
    \end{bmatrix}(qZ|q\tau),
\end{multline}
with $\bar s=0,\ldots,q-1$. Up to an overall constant
the result is identical to that obtained in Ref. \onlinecite{haldanewfs}.

\subsection{The $\nu = 2/5$  state} \label{2/5}

As a first example of a second level hierarchy state, we now explicitly compute the wave function for the filling fraction $\nu=2/5$. We use the same approach as for the Laughlin wave functions: the holomorphic part of the correlator, together with the square root  of the gaussian, yields a set of basis states. We construct all linear combinations which are eigenfunctions of the single-particle magnetic translation operators $t_{1,i}^{N_s}$ and $t_{2,i}^{N_s}$. For simplicity, we set both solenoid fluxes to zero.  This yields five candidate wave functions with $T_1$ eigenvalues $e^{2\pi i n \nu}$. 

The correlator is built from vertex operators constructed from  two bosonic fields with radii $R_1^2 = 3$ and  $R_2^2 = 15$. 
Since this is a Jain state, corresponding to two completely filled CF Landau levels, 
the sets contain equal number of electrons so that $M_1=M_2=N/2$. In the hierarchy picture this amounts to having a maximally dense condensate of quasielectrons on top of the $\nu = 1/3$ parent state.

The   charge vectors are given by 
\be{cvtwofive}
 \mbf Q^{(1)} &=& (\frac{3}{\sqrt{3}},0) \\
 \mbf Q^{(2)} &=& (\frac{2}{\sqrt{3}},\frac{5}{\sqrt{15}}) \, , \nonumber
 \ee
and we get $Z^{(1)}= 3Z_1+2Z_2 $ and $Z^{(2)} = 5 Z_2$ where $Z_1$ and $Z_2$ are the CM coordinates of the two sets respectively.  
Following the strategy outlined in Section \ref{laughlin}, we ignore the derivative operator in $V_2$ and calculate the correlators of the corresponding primary operators. 
The pertinent correlation function  decomposes as
\begin{eqnarray}
&&\langle \prod_{j\in I_1}\hat V_1(z_j, \bar z_j)\prod_{a\in I_2}\hat V_2(z_a,\bar z_a)\rangle=\\
&&\langle  \prod_{j\in I_1} e^{i\frac{3}{\sqrt{3}}\varphi_1(z_j,\bar z_j)}\prod_{a\in I_2} e^{i\frac{2}{\sqrt{3}}\varphi_1(z_a,\bar z_a)}\rangle
\langle \prod_{a\in I_2} e^{i \frac{5}{\sqrt{15}}\varphi_2(z_a,\bar z_a)}\rangle  . \nonumber
\ee
The correlator involving $\varphi_1$ contributes the following holomorphic factor
\be{F1}
 \prod_{i<j\in I_1}\vartheta_1(  z_{ij} |\tau)^3\prod_{a<b\in I_2}\vartheta_1(  z_{ab} |\tau)^{\frac{4}{3}}\nonumber\\
\times\prod_{i\in I_1, a\in I_2} \vartheta_1(  z_{ia}  |\tau)^2 \mathcal{G}^{1}_{r,t,s}(Z^{(1)}) \, ,
\ee
where $\mathcal{G}^{1}_{r,t, s}(z) \equiv \mathcal H^{(1)}_{r,t,s}(z)$. This notation is chosen to be consistent with the general hierarchy, where $\mathcal H^{(n)}_{r,t,s}$ is used for the correct CM dependence at level $n$.  Whenever convenient, we will suppress the $r$ and $t$ dependence and drop the bar on $\bar s$, still keeping track of the proper range of this index. 

The second correlator contributes a factor, which can be parametrized as 
\be{F2}
 \prod_{a<b\in I_2} \vartheta_1(   z_{ab}   |\tau)^{\frac{5}{3}}\mathcal{G}_{s}^{2}( 5 Z_2),
\ee
with 
\be{confBlock2}
\mathcal{G}_s^{2}(5 Z_2)&=& \sum_k (-1)^{t k}e^{i\pi \tau 15(k+a_2)^2}e^{2\pi i(k+a_2)5 Z_2}
\ee
and $a_2=\frac{s_2}{15}+\frac{r}{2}$. The basis set of wave functions is then obtained as a product of the conformal blocks \pref{F1} and \pref{F2}. The derivatives commute with the magnetic translation operators and \eqref{eq:pbc} must be fulfilled for all divisions into sets separately. Therefore, we can consider the functions 
\be{fulltf}
\psi =\psi_{jas} \, \mathcal{H}^{(2)}(Z^{(1)},Z^{(2)}) e^{-\frac{1}{2}\sum_k y_k^2} \, ,
 \ee
 instead of the full wave function without loss of generality.
The CM part, $\mathcal{H}^{(2)}(Z^{(1)},Z^{(2)}) $, is a sum of products $\mathcal{G}^1_{s_1}(Z^{(1)})\mathcal{G}^2_{s_2}(Z^{(2)})$ and $\psi_{jas}$ denotes the Jastrow factor expected from the result on the plane. Even though $ \mathcal{H}^{(2)}(Z^{(1)},Z^{(2)})$ depends on the CM of the sets and, therefore, on both total CM and relative coordinates, we will still call it the CM function in the following sections.

Note that we choose to express the correlators in the basis defined by \pref{Firr} rather than the one obtained directly form \eqref{eq:laughlinbasis}. This will be very convenient, since by construction the basis we use already incorporates the correct boundary conditions for the first set.
As we will see, a similar thing happens for a general hierarchy state, where  a proper choice of conformal blocks will automatically impose the correct transformation properties for all but the last set of particles.

As for the Laughlin states, we proceed by demanding  $t_{\alpha,i}^{N_s}$  to be diagonal (with eigenvalues $+1$) on the  wave function. As the transformation properties of both the Jastrow factors and the gaussian are known, we focus only on the CM dependence and derive its properties under single-particle translations of $z_i\in I_\alpha$: 
\be{trans25}
\frac{\mathcal{H}^{(2)}(Z_\alpha+ 1)}{\mathcal{H}^{(2)}(Z_\alpha)}&=& e^{i\pi (N_s-\kappa_\alpha)} \\
\frac{\mathcal{H}^{(2)}(Z_\alpha+\tau )}{\mathcal{H}^{(2)}(Z_\alpha)}&=& e^{i\pi (N_s-\kappa_\alpha)}e^{-i\pi\tau 3}
 e^{-2\pi i \mbf Q^{(\alpha)} \sum_\beta   \mbf Q^{(\beta)}    Z_\beta}  . \nonumber 
\ee
We suppress the dependence on all CM coordinates but the translated one in this and the following section. 
We also introduce  $|\mbf Q^{(\alpha)}|^2=\kappa_\alpha$ as an abbreviation. Note, that $\kappa_\alpha$ is odd for all $\alpha$'s, since the vertex operators are fermionic. Thus, all electrons obey the same boundary conditions, independent of which set they are in.

Since $\mathcal{H}^{(2)}$ is a linear combination of products 
$\mathcal{G}_{s_1}^{1} \mathcal{G}_{s_2}^{2}$,  we first determine their properties under single-particle translations. As already pointed out, the former has, by construction, the correct properties for translations of particles in the first set, but not for those in  the second:
\be{transZ2}
\frac{\mathcal{G}_{s_1}^{1}(Z^{(1)}+c_1)}{\mathcal G_{s_1}^{1}(Z^{(1)})}&=&e^{2\pi i a_1 2} \nonumber\\
\frac{\mathcal{G}_{s_1}^{1}(Z^{(1)}+ c_1\tau )}{\mathcal G_{s_1+2}
^{1}(Z^{(1)})}&=& e^{-2\pi i(2Z_1+\frac{4}{3}Z_2)}e^{-i\pi \tau \frac{4}{3}} \, .
\ee
Comparing these expressions with the ones of $\mathcal{G}_{s_2}^2(Z^{(2)})$:
\be{transZ2F}
\frac{\mathcal{G}_{s_2}^2(Z^{(2)}+ q_2)}{\mathcal{G}_{s_2}^2(Z^{(2)})}&=&e^{2\pi i a_2 5} \nonumber\\
\frac{\mathcal{G}_{s_2}^{2}(Z^{(2)}+ q_2\tau)}{\mathcal{G}_{s_2+5}^{2}(Z^{(2)})}&=& e^{-2\pi i \frac{5}{3}Z_2}e^{-i\pi \tau \frac{25}{15}} \, ,
\ee
we find that the coordinate dependent factors (independent of the free parameters $s_1$ and $s_2$) combine to give $2Z_1+3Z_2=\mbf Q^{(2)}\cdot \sum_\alpha \mbf Q^{(\alpha)}\ Z_\alpha$ and, thus, the correct factor to cancel the one from the relative part and the gaussian. This is not a coincidence, as it may seem here, but  a general property due to the construction of the vertex operators. Correct transformation properties along $L_1$ require special combinations of $s_1$ and $s_2$, so that $(2 s_1+s_2)/3$ is an integer. Hence, we find that only the combinations where $-s_1+s_2=0$ mod 3 are consistent with the eigenvalues in \pref{trans25}. This reduces the number of 'good' basis states from 45 to 15 ($\times$ 4 for the different flux sectors). By taking $ r=(N_s-\kappa_1)$ mod $2$  we obtain the correct sign in \pref{trans25}. 
 
We can now use this reduced set to find the linear combinations that also transform correctly under $t_{2,i}^{N_s}$. It is easy to see that if a pair $(s_1 ,s_2)$ satisfies $-s_1+s_2= 3k$, then $(s_1+2, s_2+5)$ does, too. Therefore, $t_{2,i}^{N_s}$ only maps functions in this reduced set into each other.  Hence, we use the parametriztion  $(s_1 ,s_2) = (2l, 5l+3s')$ and form the  linear combination 
\be{cb25}
\mathcal H_{\bar s}^{(2)}(Z^{(1)},Z^{(2)})&=&\sum_{l=0}^{2}(-1)^{tl} \mathcal G_{2j}^{1}(Z^{(1)},Z^{(2)})\nonumber\\
&\times& \mathcal G_{5l+3 s'}^{2}(Z^{(2)}) \, ,
\ee
which transforms correctly, if $t$ is chosen to be: $t=N_s-\kappa_1$. Note, that we use $\bar s=3 s'$ mod 5, $s'=0,\ldots,4$ to label the CM coordinate. That this is a natural choice is seen by the action of $T_1$ on (\ref{cb25}), 
\be{25T1}
T_1\mathcal H_{\bar s}^{(2)}(Z^{(1)},Z^{(2)})&=&\sum_{l=0}^{2}(-1)^{tl} \mathcal G^1_{2l}(Z^{(1)})\mathcal G^2_{5l+3s'}(Z^{(2)})\nonumber\\
&\times& e^{2\pi i a_1(3+2)\frac{N}{2N_s}}e^{2\pi i a_2 5\frac{N}{2N_s}}\nonumber\\
&=&  \mathcal H_{\bar s}^{(2)}(Z^{(1)},Z^{(2)}) e^{2\pi i \frac{2}{5} \bar s} \, ,
\ee
where $\bar s $ naturally occurs multiplied with the filling fraction $\nu = 2/5$, which amounts to $K_1=(2 N \bar s)\mod N_2$. 
In addition, $T_2$ acting on $\mathcal H_{\bar s}^{(2)}$ only shifts $\bar s\rightarrow \bar s+1$. Thus, $\mathcal{H}_{\bar s}^{(2)}$ is invariant under $T_2^5$ as expected and $K_2=0$.  There is no overall sign, as both $\mathcal G^1$ and $\mathcal G^2$ pick up a factor $(-1)^t$ under $\bar s\rightarrow \bar s+5$.

In order to get the wave function, we need to reintroduce the derivatives and antisymmetrize over all possible divisions into the two sets:
\be{wf25}
\Psi_{2/5} &=& \sum_{i_1<i_2<...i_{N/2}\atop a_1<a_2<...a_{N/2}}(-1)^{\sum_j a_j} \prod_{k} \mathcal{D}_{a_k}^{(1)} \prod_{i_j<i_l}\vartheta_1(z_{i_ji_l}|\tau)^3\nonumber\\
&\times&\prod_{a_j<a_l}\vartheta_1(z_{a_ja_l}|\tau)^3\prod_{i_j, a_l}\vartheta_1(z_{i_ja_l}|\tau)^2 \nonumber\\
&\times&\mathcal{H}_{\bar s}^{(2)}(Z^{(1)},Z^{(2)}) e^{-\frac{1}{2\ell^2}\sum_k y_k^2}
\ee
with $Z^{(1)}=3\sum_{j}z_{i_j}/L_1+2\sum_j z_{a_j}/L_1$ and $Z^{(2)}=5\sum_j z_{a_j}/L_1$. 
There are $\lceil {N_s} / {2}-1\rceil$ possible choices for the derivative operator for a fixed set of quantum numbers of $\Psi_{2/5}$. 
 More comments on this  can be found in Section \ref{results} where we test the candidate wave function numerically against exact diagonalization results.

\subsection{The general hierarchy state}\label{sechierarchy}
In this section we generalize the previous construction of the Laughlin  and $\nu = 2/5$ wave functions on the torus to the general class of filling fractions discussed in Section \ref{hwf}. 

Recall that to construct the hierarchy states at level $n$ the electrons are divided into $n$ sets $I_\alpha$. 
The electrons in set $I_\alpha$ are represented by $V_{\mbf Q^{(\alpha)}}$, giving a center of mass coordinate $Z_\alpha=\sum_{k\in I_\alpha}z_k/L_1$. 
Thus, the correlator we want to compute is of the form:
 \be{corr}
\langle\prod_{\alpha=1}^n \prod_{k\in I_\alpha}\hat V_{\mbf Q^{(\alpha)}}(z_k,\bar z_k)\rangle \, .
\ee
It factorizes into a product of correlators containing only one field, each of which can be calculated as in Section \ref{wfs}. 
In the following, we use the same conventions as in the previous section. We denote the functions $\mathcal G^a_{r,t,s_a}$  as CM function, even though they depend on the CM of the sets and not the total CM.  For simplicity, the fluxes are set to zero, $\phi_1=\phi_2=0$, and we suppress the indices $r$ and $t$ as they are fixed by the boundary conditions. 

The wave functions are extracted from (\ref{corr}) by exactly the same approach as used in the Laughlin case and $\nu=2/5$. As in the latter, the wave function is an antisymmetric sum over the various ways to divide the electrons into the sets. Each summand must be an eigenfunction of the magnetic translation operators. This constraint is used to construct the suitable linear combinations of the conformal blocks for any given division. The basis states are given by
\be{basis states}
\psi_{r,t,\{s_a\}} = \prod_{\alpha=1}^n\prod_{i_\alpha< j_\alpha \in I_\alpha}\vartheta_1(z_{i_\alpha j_\alpha}|\tau)^{\mbf Q^{(\alpha)} \cdot  \mbf Q^{(\alpha)}} \prod_{\alpha < \beta}
\nonumber\\
\times\!\prod_{i_\alpha\in I_\alpha\atop j_\beta\in I_\beta}\vartheta_1(z_{i_\alpha j_\beta}|\tau)^{\mbf Q^{(\alpha)}\cdot \mbf Q^{(\beta)}} \prod_{a=1}^n \mathcal{G}_{s_a}^{a}(Z_n^{(a)})e^{-\frac{1}{2}\sum_k y_k^2}\, ,
\ee
where we used the same conventions for $\mathcal G^j_{s_j}$ as in the $2/5$ case:
\begin{multline}\label{holblock}
\mathcal{G}_{s_j}^{j}(Z_n^{(j)})= \sum_k (-1)^{t k} e^{i\pi \tau R_j^2(k+a_j)^2} e^{2\pi i (k+a_j)Z_n^{(j)}}\, ,
\end{multline}
with $a_j= \frac{s_j}{ R_j^2}+\frac{r}{2}$ and $s_j=1,\ldots 4R_a^2-1$. Their transformation under translations along $L_1$ and $L_2$ follow directly from \eqref{eq:F_transf}:
\be{GtransZi}
\frac{\mathcal{G}_{s_j}^j(Z_n^{(j)}+c)}{\mathcal{G}_{s_j}^j(Z_n^{(j)})}&=&e^{2\pi i a_j c} \nonumber\\
\frac{\mathcal{G}_{s_j}^j(Z_n^{(j)}+c\tau)}{\mathcal{G}_{s_j+c}^j(Z_n^{(j)})}&=& e^{-2\pi i \frac{c}{R_j^2}Z_n^{(j)}}e^{-i\pi \tau \frac{c^2}{R_j^2}} \, ,
\ee
and they have the periodicity property
\be{gper}
\mathcal{G}_{s_j+ R_j^2}^{j}(Z_n^{(j)})= 
(-1)^t e^{-i\pi \tau R_j^2}e^{-2\pi i Z_n^{(j)}}\mathcal{G}_{s_j}^{j}(Z_n^{(j)}) \, .
\ee
The functions $\mathcal{G}^j$ depend only on  $Z_n^{(j)}=c_j\sum_{\alpha=j+1}^n Z_\alpha +q_j Z_j
$ and are thus independent of the first $j-1$ CM coordinates. 
In addition, they are identical for the parent and daughter state, except that $Z_n^{(j)}$ must be replaced by $Z_{n+1}^{(j)}$. This yields a recursive construction of the full CM dependence $\mathcal{H}^{(n)}_{s_n}$. To be more specific, we will show by recursion that at filling fraction $\nu_n={p_n}/{q_n}$ there are exactly $q_n$ possible CM functions $\mathcal{H}^{(n)}_{\bar s_n}$ given by:
\be{CMresult}
\mathcal{H}^{(n)}_{\bar s_n}(Z_n^{(1)},..., Z_n^{(n)}) \\
=\sum_{l_2=0}^{q_1-1}\ldots \sum_{l_{n}=0}^{q_{n-1}-1} \prod_{a=1}^{n}&& (-1)^{t (l_2+\ldots + l_{n})}\mathcal{G}_{s_a}^a(Z_n^{(a)}) , \nonumber
\ee
where
 \be{sumvar} 
 s_a&=&q_a l_a+\sum_{k=a+1}^n c_a l_k \hspace{0.4cm} ,\hspace{0.5cm}   a<n  \nonumber\\
 s_n&=&q_n l_n+q_{n-1}s\, \, .
 \ee  
 In analogy with the $\nu = 2/5$ case, and for reasons that will become clear below, we choose $\bar s_n$  as the free parameter, with $\bar s_n = q_{n-1}s$ mod $q_n$, $s=0,\ldots,q_n-1$. 

We have already shown that the CM part of the Laughlin wave functions is of the form (\ref{CMresult}). It remains to show that, assuming (\ref{CMresult}) for a parent state at level $n$, it follows that also the CM part of the daughter state at level $n+1$ obeys this relation.   
To go from level $n$ to level $n+1$ in the hierarchy, we consider a quasiparticle density of ${1}/{(2k_{j+1}+1})$ atop the parent state. The daughter state is then at filling fraction $\nu_{n+1}$  determined by \eqref{fillfr}. To construct the wave function, there is then an additional vertex operator,$V_{n+1}$, with charge vector $\mbf Q^{(n+1)}=(\frac{c_1}{R_1},...,\frac{c_n}{R_n}, \frac{q_{n+1}}{R_{n+1}})$. Note, that all charge vectors are now $n+1$ dimensional objects.
The wave function will be of the form
\be{wfhier}
\psi&=& \psi_{jas}(z_{ij}) \mathcal{H}^{(n+1)}_{\bar s_{n+1}}(Z_{n+1}^{(1)},...,Z_{n+1}^{(n+1)})e^{-\frac{1}{2}\sum_i y_i^2},
\ee
where $\mathcal{H}^{(n+1)}_{\bar s_{n+1}}$ is some suitable linear combination of products of $\mathcal{G}^a_{s_a}$. 

Invariance under one-particle magnetic translations puts constraints on the wave function. As the Jastrow type factor is known, we can compute which relations  $\mathcal{H}^{(n+1)}_{\bar s_{n+1}}$ must satisfy under one-particle (normal) translations.  Under the translations $Z_\alpha\rightarrow Z_\alpha+1$ and $Z_\alpha\rightarrow Z_\alpha+\tau$, $\mathcal{H}^{(n+1)}_{\bar{s}_{n+1}}$ transforms as
\be{transprop1}
\frac{\mathcal H_{\bar{s}_{n+1}}^{(n+1)}(Z_\alpha+1)}{\mathcal H_{\bar s_{n+1}}^{(n+1)}(Z_\alpha)}&=& (-1)^{(N_s-\kappa_\alpha)}\nonumber\\
\frac{\mathcal H_{\bar s_{n+1}}^{(n+1)}(Z_\alpha+\tau)}{\mathcal H_{\bar s_{n+1}}^{(n+1)}(Z_\alpha)}&=& (-1)^{(N_s-\kappa_\alpha)} e^{-2\pi i \mbf Q^{(\alpha)}\cdot \sum_\beta \mbf Q^{(\beta)}Z_\beta}\nonumber\\
&\times&e^{-i\pi\tau \kappa_\alpha},
\ee
where we use the same simplified notation as in \eqref{trans25}: $\mathcal H^{(n+1)}_{\bar s_{n+1}}(Z_\alpha+c)=\mathcal H^{(n+1)}_{\bar s_{n+1}}(Z_{n+1}^{(1)},...,Z_{n+1}^{(n+1)})$, but with $Z_\alpha$ replaced by 
$Z_\alpha+c$ in the $Z^{(i)}$'s.
In deriving the second of these relations, we used \pref{Ns-rel}. Equation \eqref{transprop1} is valid for the CM dependent functions at all levels, in particular also at level $n$.

The function $\mathcal{H}^{(n+1)}_{\bar s_{n+1}}$ is a linear combination of products: $\prod_{j=1}^{n+1} \mathcal{G}^{(j)}_{s_j}(Z_{n+1}^{(j)})$.  The task of finding the correct linear combination is simplified by using the result for the parent state. By construction, $\mathcal{H}_{\bar s_{n}}^{(n)}(Z_{n+1}^{(1)},...,Z_{n+1}^{(n)})$ transforms correctly under  translations of all CM coordinates $Z_\alpha$ but $Z_{n+1}$ . For this last coordinate, note that for $a<n$, $\mathcal{G}^{(a)}$ depends in the same way on $Z_{n+1}$ as on $Z_{n}$, as $Z^{(a)}_{n+1}=q_a Z_a+c_a(Z_{a+1}+...+Z_n+Z_{n+1})$. This is not true for $\mathcal G^{(n)}_{s_n}$, but we can separate the difference by writing:
\be{hint}
\mathcal{G}^{(n)}_{s_n}(Z_{n+1}^{(n)})
&=&\mathcal{G}_{s_n}^{(n)}(q_n (Z_n +  Z_{n+1})) f(Z_n,Z_{n+1})
\ee
with
\be{ratio}
f(Z_n,Z_{n+1})&=&\frac {\mathcal{G}_{s_n}^{n}(q_n Z_n + c_n Z_{n+1})}{\mathcal{G}_{s_n}^{n}(q_n Z_n + q_n Z_{n+1})}  \, .
\ee
This implies that for a given set $(l_2, \dots ,\l_n)$ in \pref{CMresult}, the product $\mathcal{G}^1\cdot...\cdot\mathcal{G}^{n}$ transforms  in the same way under translations of $Z_n$ and $Z_{n+1}$, except for the ratio \eqref{ratio}. The transformations of this ratio is, however, readily obtained from \pref{holblock}, and since it is independent of $l_n$, we can infer the following relation for translations along $L_1$: 
\be{transpropL1}
\mathcal{H}_{\bar s_n}^{(n)}(Z_{n+1}+1)&=&
e^{-2\pi i  \frac{\bar s}{q_n}}\mathcal{H}_{\bar s_n}^{(n)}(Z_{n+1}) \, ,
\ee
and a slightly more complicated relation for translations along $L_2$:
\be{transpropL2}
&&\mathcal{H}_{\bar s_n}^{(n)}(Z_{n+1}+\tau)\nonumber\\
&&=e^{-2\pi i\sum_{\beta=1}^{n}(\mbf Q^{(n+1)}\cdot \mbf Q^{(\beta)}Z_\beta+\frac{c_{\beta}^2}{R_\beta^2}Z_{n+1})}
e^{-i\pi\tau(\kappa_n+\frac{c_n^2-q_n^2}{R_n^2})}\nonumber\\
&&\times\sum_{l_2=0}^{q_1-1}\ldots \sum_{l_{n}=0}^{q_{n-1}-1} \prod_{a=1}^{n} (-1)^{t (l_2+\ldots + l_{n})}\mathcal{G}_{s_a+c_a}^a(Z_{n+1}^{(a)}) \nonumber\\
&&=\pm e^{-2\pi i\sum_{\beta=1}^{n}(\mbf Q^{(n+1)}\cdot \mbf Q^{(\beta)}Z_\beta+\frac{c_{\beta}^2}{R_\beta^2}Z_{n+1})}
e^{-i\pi\tau(\kappa_n+\frac{c_n^2-q_n^2}{R_n^2})}\nonumber\\
&&\times \mathcal{H}_{\bar s_n+c_n}^{(n)}(Z_{n+1}) \, .
\ee
The sign is calculable, but is not  needed in the following discussion.  As $q_n$ and $q_{n+1}$ are relatively prime, $j c_n$ mod $q_n$ runs through all possible values of $q_n$. Therefore, invariance under translations of the set $I_{n+1}$ will require linear combinations of all $q_n$ CM functions of level $n$.

To obtain an eigenfunction of $t_{\alpha i}^{N_s}$, $\mathcal{H}^{(n)}_{\bar s_n}$ must be combined with $\mathcal{G}_{s_{n+1}}^{n+1}(Z_{n+1}^{(n+1)})$  which transforms according to  \eqref{GtransZi} with $c=q_{n+1}$:
\be{transpropF}
\frac{\mathcal{G}_{s_{n+1}}^{n+1}(Z_{n+1}^{(n+1)}+q_{n+1})}{\mathcal{G}_{s_{n+1}}^{n+1}(Z_{n+1}^{(n+1)})}&=&
e^{2\pi i \frac{s_{n+1} }{q_{n}}}\nonumber\\  \\
\frac{\mathcal{G}_{s_{n+1}}^{n+1}(Z_{n+1}^{(n+1)}+q_{n+1}\tau)}{\mathcal{G}_{s_{n+1}+q_{n+1}}^{n+1}(Z_{n+1}^{(n+1)})}&=&
e^{-i\pi\tau\frac{q_{n+1}^2}{R_{n+1}^2}}e^{-2\pi i\frac{q_{n+1}^2}{R_{n+1}^2}Z_{n+1}}  \, . \nonumber 
\ee
Comparing the first lines of (\ref{transprop1}), (\ref{transpropL1}) and \pref{transpropF} we see that the condition
\be{cond}
-\bar s_n+s_{n+1}= m \, q_n \ \ \ \ ; \ \ \ m \ \ \   {\rm integer}
\ee must be fulfilled. This reduces the number of allowed products of conformal blocks from $q_nR_{n+1}^2$ to $R^2_{n+1}$. A further reduction is obtained from requiring invariance under translations along the $\tau$-direction. 
The only combinations that transform correctly are given by
\be{resultn+1}
&&\mathcal{H}^{(n+1)}_{\bar{s}_{n+1}}(Z_{n+1}^{(1)},\ldots, Z_{n+1}^{(n+1)})\\
&&=\sum_{j=0}^{q_n=1} (-1)^{tj}\mathcal{H}^{(n)}_0(Z_{n+1}+j\tau)\mathcal{G}^{n+1}_{q_n s}(Z_{n+1}+j\tau), \nonumber
\ee
where, again, $\bar{s}_{n+1}=q_ns$ and  $s=0,\ldots,q_{n+1}-1$. Inserting \eqref{transpropL2} into the above equation, it is straightforward to verify that it is of the form \eqref{CMresult}. We chose $\bar s_n=0$ in \eqref{resultn+1} in accordance with the sign convention in \eqref{CMresult}. Taking another value amounts only to an overall sign change. 

We can also compute the quantum numbers of  the wave functions \eqref{wfhier} under magnetic translations recursively. Under translations with $T_1$, the conformal blocks at level $n+1$ pick up a phase given by: $(-1)^{r(n+1)} e^{2\pi i \nu_{n+1} \bar s_{n+1}}$. Thus, $K_1$ is given by  $K_1=(r (n+1) N_s/2+p_{n+1} N \bar s_{n+1}) $ mod $N_s$. We can also show that $K_2=t(n+1) N_s/2$ mod $N_s$. As in the previous cases, the different fluxes $\phi_\mu$  in \pref{eq:pbc} can be incorporated  by a proper choice of $r$ and $t$. 

Hence, we found exactly $q_{n+1}$ conformal blocks with the correct quantum numbers. The reader may also note here that the total number of conformal blocks (before imposing the boundary conditions) is not given by the filling factor. Other choices for the charge vectors may decrease this number considerably and simplify computations. This will be discussed  in more detail in the next section.

In conclusion, we have constructed the hierarchy wave functions in torus geometry for all filling fractions $\nu_n$ that are obtained by successive condensation of quasielectrons. At level $n$ we find $q_n$ wave functions that are eigenfunctions of the magnetic translation operators $T_1$ and $T_2^{q_n}$, confirming the expected ground state degeneracy on the torus. Reinstalling the derivatives and antisymmetrizing, we obtain an explicit expression for the wave functions at filling factor $\nu_n$:

\be{final}
\Psi_{\nu_n,\bar s_n}&=&\sum_{I_1,...,I_n} (-1)^\sigma \prod_{k=2}^n \prod_{i_k\in I_k} \mathcal D^{(k-1)}_{i_k} \prod_{\alpha<\beta}\prod_{i_\alpha\in I_\alpha\atop j_\beta\in I_\beta}\nonumber\\
&\times&\vartheta_1(z_{i_\alpha j_\beta}|\tau)^{\mbf Q^{(\alpha)}\cdot\mbf Q^{(\beta)}} \prod_{\alpha=1\atop i_\alpha< j_\alpha \in I_\alpha}^n\vartheta_1(z_{i_\alpha j_\alpha}|\tau)^
{\kappa_\alpha}\nonumber\\
&\times&\mathcal{H}_{\bar s_n}^{(n)}(Z_n^{(1)},..,Z_n^{(n)}) e^{-\frac{1}{2}\sum_k^N y_k^2}.
\ee
The sum runs over all possible ways to divide $N$ particles into $n$ sets of size $M_1,\ldots, M_n$ and $(-1)^\sigma$ is the sign picked up by rearranging the radially ordered particles into the sets.

\section{Alternative charge vectors}\label{sym}

We have already mentioned that the charge vectors $\mbf Q^{(\alpha)}$ are not uniquely determined by the wave function. To spell out this ambiguity we first notice from \pref{basis states} that the charge vectors enter in the relative part only through the scalar products $\mbf Q^{(\alpha)}\cdot\mbf Q^{(\beta)}$. If we introduce the vector $c_\alpha = 1/R_\alpha$, the holomorphic current operator takes the form: $J(z) =i\sum_\alpha c_\alpha \partial \varphi_\alpha (z)$. The constraints on the vertex operators that are imposed by the short-distance behaviour of the electrons, unit $U(1)$ electric charge and the filling fraction,  can be expressed as the following conditions on  the $n$-dimensional vectors $\mbf Q^{(\alpha)}$ and $\mbf c$,
\be{ccond}
 \mbf Q^{(\alpha)}\cdot\mbf Q^{(\beta)} &=& K^{\alpha\beta}\nonumber \\
\mbf  Q^{(\alpha)} \cdot \mbf c &=&1 \nonumber \\
\mbf c \cdot \mbf c &=& \nu \, .  
\ee
These scalar products are invariant under simultaneous $O(n)$ transformations on $\mbf Q^{(\alpha)}$ and $\mbf c$. (It should also be clear how to include quasiholes by introducing a new set of vectors, $\mbf l^{(\alpha)}$.) We note, however, that an $O(n)$ transformation will in general give irrational radii implying that the CFT is not rational. This means that the charge sums cannot in any obvious way be reorganized as a finite sum over conformal blocks as required by the  general analysis in Section \ref{wfs}. Naively it would seem that  no similar restrictions to rational radii would apply when working in the plane, but it is not unlikely  that they would emerge from a study of the edge theory for a finite droplet. 

We have not tried to find the most general transformation that leaves \pref{ccond} invariant and still maintains rational radii of the fields, but only studied a few examples. Note that it is not obvious that preserving \pref{ccond} automatically implies that the CM part of the wave function remains the same. On the contrary, this amounts
to non-trivial mathematical relations between conformal blocks on rational tori. We have numerically verified such a relation explicitly in the case of $\nu=2/5$ and $\nu=3/7$, and suggest that similar results hold in general.  
If, on the contrary, these relations turned out not to hold, it would indicate the presence of inequivalent hierarchy states at the same level, and with the same filling fraction.

We now present alternative charge vectors for the fractions 2/5, 4/11 and 3/7, in a basis where $\mbf c = (\sqrt \nu,0, \dots ,0)$. In this basis a background charge is needed only for the field $\varphi_1$, since the correlators for the remaining fields are neutral by construction. As we shall see, the number of conformal blocks is much smaller in this basis, which simplifies calculations. On the other hand, the hierarchy structure is not manifest, \eg the computation of $\nu=2/5$ blocks is in no simple way  related to the  computation at  $\nu=1/3$.

For $\nu=2/5$, the charge vectors in this basis
are given by the following symmetric expressions,
\begin{align}\label{alt}
  \mbf Q^{(1)} &= ( \frac{5}{\sqrt{10}},\frac{1}{\sqrt{2}})\nonumber \\
  \mbf Q^{(2)} &= ( \frac{5}{\sqrt{10}},-\frac{1}{\sqrt{2}}) \, ,
\end{align}
where the sets $I_1$ and $I_2$ each contain $N/2$ electrons.
Since the radii squared are even integers, 
the double charge-lattice sums reduce to single sums ($p_i'=1$),
and the conformal correlator gives a holomorphic basis
of the form 
\begin{multline}
  \psi_{s,s'}(z_i) = \prod_{i<j\in I_1}\vartheta_1(z_{ij}|\tau)^3
  \prod_{a<b\in I_2}\vartheta_1(z_{ab}|\tau)^3\\
\times  \prod_{i\in I_1, a\in I_2}\vartheta_1(z_{ia}|\tau)^2
  F_s^{10}(5Z)F_{r}^2(Z_{12}) \, ,
\end{multline}
where $Z = Z_1+Z_2$, $Z_{12}=Z_1-Z_2$, and
\begin{align}\label{Fsym}
  F_s^{10}(5Z) &= \sum_k e^{i\pi
\tau10(k+s/10)^2}e^{2\pi i(k+s/10)5Z}\nonumber\\
  F_{s'}^2(Z_{12}) &= \sum_k e^{i\pi \tau 2(k+s'/2)^2}e^{2\pi i (k+s'/2)Z_{12}}.
\end{align}
The first conformal block depends on the total CM, while the second contains only relative coordinates. However, it is still not possible to separate the CM dependence, as explained below. 
A linearly independent basis of 20 states is obtained by taking $s=0,\ldots, 9$, $s'=0,1$. 
This should be compared with the calculation in Section IVB, where the original 720 states
had to be reduced to 20 by imposing proper boundary conditions. 

To compare with our previous calculation we change the basis for the $\varphi_1$ field  to
\begin{align}\label{F10alt}
\tilde F^{10}_{s,t}(5Z)&= \!\sum_k e^{i\pi
\tau10(k+s/5+t/2)^2}e^{2\pi i(k+s/5+t/2)5Z} \, ,
\end{align}
where the parameters have the values $s=0,\dots,4$ and $t=0,1$. In this parametrization, $s$  labels the CM coordinate (in fact, this is precisely $s'$ in section \ref{2/5}), while the boundary conditions are coded in $r$ and the different combinations of $t$ and $s'$. 
An analysis along the lines of that given in the previous examples, shows that
the following linear combinations give eigenstates of $t_{1,i}^{N_s}$ and $t_{2,i}^{N_s}$:

\begin{eqnarray}
\mathcal{H}^{(2)}_s(Z_1,Z_2)&=& \tilde F_{s,0}^{10}(5Z)F_\alpha^2(Z_{12})\nonumber\\
&+&(-1)^\beta \tilde F_{s,1}^{10}(5Z) F_{\alpha+1}^2(Z_{12}) \, ,
\ee
where we choose $\alpha=(N_s-\kappa_1+\phi_1/\pi)$ mod 2 and $\beta=(N_s-\kappa_1+\phi_2/\pi)$ mod 2. Note, that the boundary conditions require a non-trivial combination of different CM functions and, thus, there is no simple way to factor out the total CM. As already mentioned, we have checked that the wave functions thus obtained are numerically equal to those given in \pref{cb25}.  This amounts to a rather complicated identity between sums of products of generalized theta functions. 

Finally, we also give explicit expressions for the charge vectors for $\nu=4/11$ and $\nu=3/7$. In the former, we have different number of particles in the two sets and, thus, the charge vectors look less symmetric: 
\begin{equation}
{\mbf Q}^{(1)}=(\sqrt{\frac{11}{4}},\frac{1}{2}),\hspace{1cm}
{\mbf Q}^{(2)}=(\sqrt{\frac{11}{4}},-\frac{3}{2}) \, .
\end{equation} 
For $3/7$ we find the three charge vectors: 
\begin{eqnarray}
{\mbf Q}^{(1)}&=&(\sqrt{\frac{7}{3}},\frac{2}{\sqrt{6}},0)\nonumber\\
{\mbf Q}^{(2)}&=&(\sqrt{\frac{7}{3}},-\frac{1}{\sqrt{6}},\frac{1}{\sqrt 2}) \nonumber\\
{\mbf Q}^{(3)}&=&(\sqrt{\frac{7}{3}},-\frac{1}{\sqrt{6}},-\frac{1}{\sqrt 2})\,\, .
\end{eqnarray}
Using these, the number of conformal blocks is greatly reduced compared with the
earlier representation \pref{chargevector}, but unlike the $\nu=2/5$ case, there remain states that must be excluded by applying the boundary conditions.

\section{Numerical tests}\label{results}

We have carried out numerical tests of some of the simplest hierarchy wave functions, namely the $\nu = 2/5$ Jain state, and the recently discovered $\nu =4/11$ state, both at level two, and the level three Jain state at $\nu = 3/7$. Here we present results for the 2/5 state and just comment briefly on the others at the end of the section. 
We compare our wave functions to the ground states obtained by exact diagonalization of eight respectively ten particles using an unscreened Coulomb interaction. 

As discussed in Section \ref{bgcharge}, the wave functions we have constructed are not unique because of the 
freedom associated with the derivative operators ${\cal D}^{(\alpha )}$. In particular, for the 2/5 state we have to define the operator ${\cal D}^{(1)} $. A simple set of choices is,
\be{choice}
{\cal D}^{(1)}_n = t_1^n
\ee
and we shall denote the corresponding wave functions by  $\psi^{(n)}_{2/5}$. For the simplest choice of $n=1$, 
 we find that the overlap between the exact groundstate and $\psi_{2/5}^{(1)}$ is well above $0.9$ for all values of $L_1$ and $N\leq 10$. For some values of $L_1$ this overlap even exceeds $0.99$, see the lowest lying curves in Figs  \ref{fig:projN8} and \ref{fig:projN10}. 
 
The results are improved further if we take a linear combination of states $\psi^{(n)}_{2/5}$ with different values of $n$.  We find that the overlap with the exact states and the space spanned by $\{\psi_{2/5}^{(n)}\}_{n=1}^{k}$ for different values of $k$  quickly converges to a  number very close to one---in fact, above 0.99 already when two or three states are taken into account---irrespective of the values of $L_1$ and $N\leq 10$. These results for eight and ten electrons are summarized in Figs \ref{fig:projN8} and \ref{fig:projN10}.  We note that the small variation of the overlap as a function of $L_1$ very much resembles results obtained earlier for the Laughlin state on the torus \cite{yang}. 

We find the fact that we only need to take two or three states into account to get a wave function as good as the Laughlin state as strong evidence for the correctness of our approach, especially in the light of that the Hilbert space in the $\mathbf{K}=\mathbf{0}$ sector  for ten particles at $\nu=2/5$ has 26152 dimensions.

\begin{figure}[h!]
  \includegraphics[width=.98\columnwidth]{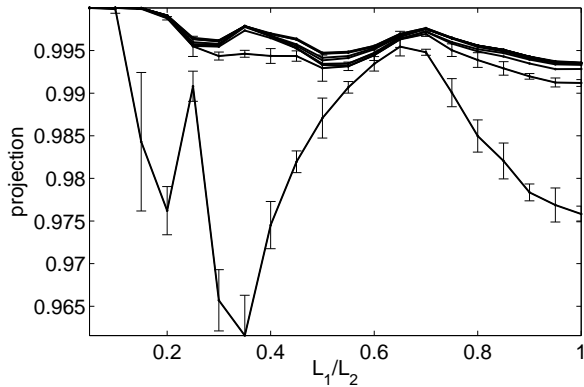}
  \caption{
    Projection (amplitude) of the eight-particle exact solution (obtained by diagonalization of unscreened Coulomb interaction)  to the subspace spanned by
    $\{\psi_{2/5}^{(n)}\}_{n=1}^{k}$, with $k=1,\ldots, 9$.
  } \label{fig:projN8}
\end{figure}

\begin{figure}[h!]
  \includegraphics[width=.98\columnwidth]{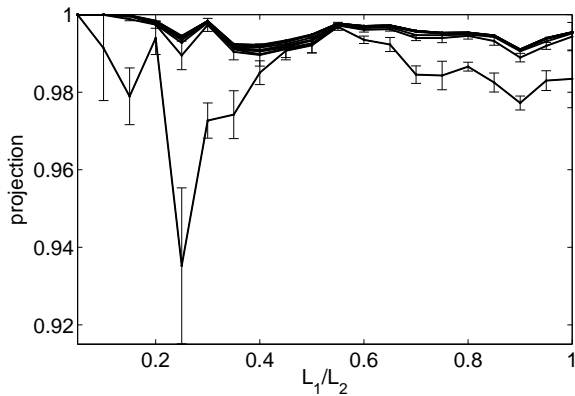}
  \caption{
    Projection of the ten-particle exact solution to the subspace spanned by
    $\{\psi_{2/5}^{(n)}\}_{n=1}^{k}$, with $k=1,\ldots, 12$. } \label{fig:projN10}
\end{figure}

Preliminary results for the state at $\nu=4/11$---one of the newly observed non-Jain states---also seems very promising. However, 
both the numerical studies and the interpretation thereof are more involved and will be published elsewhere.
Moreover, we have compared the trial wave function at filling fraction $\nu=3/7$ with exact diagonalization results and found that already the simplest choice of the derivative operators \eqref{deran} gives a reasonably good description.

\section{Summary and outlook}\label{concl}

To summarize, we have constructed torus versions of newly proposed wave functions 
describing the Haldane-Halperin hierarchy of incompressible quantum Hall states in the lowest Landau level. 
In particular, we managed to incorporate the homogeneous background charge, and derive the non-holomorphic gaussian factor, in a mathematically sound manner. In the previous calculations in the disk geometry boundary terms had to be ignored \cite{hans}, and in a spherical geometry one typically put a compensating charge at infinity at the price of not obtaining the gaussian factor.

At a technical level we note that all the boson radii $R_a$ are square roots of integers. One can construct wave functions based on CFT where (some of) the $R_a$'s are rational, but these states do not seem to be part of the usual hierarchy, and might even be non-abelian.

We believe that the  techniques developed in this paper should be useful also to construct other QH states, as, for example,  the Halperin $(n_1,n_2,m)$ states. 
It should also be possible to construct quasiparticle  states using similar techniques. For the quasiholes in the hierarchy this should be straightforward, since they are described by local vortex operators that are primary fields in the rational CFT's used in our construction. The quasielectrons are more difficult, but a suitable adaptation of the methods developed in \cite{hans} is likely to work. 
We also believe that the quasielectron excitations of  the Moore-Read pfaffian state can be obtained.

Finally we note that our construction provides a way of investigating the adiabatic continuity from the solvable thin torus case to the bulk for the hierarchy states considered here.

{\flushleft{\bf Acknowledgements}}\\
We thank  Eddy Ardonne for many helpful discussions and comments on the manuscript.
We also thank Fawad Hassan for teaching us about conformal field theory. 
This work was supported by the Swedish Research Council and NordForsk.
J.S. thanks Ari Harju and Risto Nieminen for initiating this collaboration and acknowledges the support of Vilho, Yrj\"o, and Kalle V\"ais\"al\"a Foundation.

\appendix

\section{The background charge contribution}\label{app:bg}

In this Appendix we evaluate the integrals in \pref{part2} and \pref{part3} that originate from the
homogeneous background charge. We  consider a general  rectangular integration domain defined by 
the complex number $\xi = a + ib $
\be{intdom} 
\int d^2 x\, \equiv \int_{a}^{a+L_1}dx \, \int_{b}^{b+L_2}dy \, .
\ee
The integral in \pref{part2} is now easily evaluated, and, when combined with the second exponential
yield factors,
\begin{align}\label{typterm} 
\exp \left[ \frac {2\pi i} {R_a}\left\{  c_a \sum_{\alpha = a+1}^{n} Z_\alpha + q_a  Z_a 
 - N_s\left(\frac{\xi}{L_1}-\frac{1+\tau}{2}\right)  \right\}  \right].
\end{align}
First notice that for the choice $\xi =(L_1+i L_2)/2$, corresponding to the symmetric integration region used in the main text,
the integral in \pref{part2} vanishes. For an arbitrary $\xi$ the effect of the integral can be incorporated by shifting the 
coordinates $z_i \rightarrow z_i + (\frac{\xi}{L_1}-\frac{1+\tau}{2})$ provided 
\begin{align}\label{homcons}
\forall a:\ \ \ \sum_{\alpha=1}^n Q^{(\alpha)}_aM_\alpha=C_a N_s.
\end{align}
These relations, which determine the $M_\alpha$'s, were derived previously in the plane as a consequence of the assumption 
of homogeneity. It is to be noted that exactly the same relations are needed on the torus to guarantee that the CM part of the wave function factorizes. Without starting from the plane we could thus have obtained \pref{homcons} as a consistency condition directly on the torus. Also note that since \pref{homcons} determines the $M_\alpha$'s, it also implies the relation \pref{Ns-rel} that was crucial for obtaining explicit expressions for the CM dependence for the hierarchy states. 
We now turn to the integral in \pref{part3}
\begin{equation}
  \label{eq:Aintegral}
  I_{\xi}(z,\bar z) = \int \rmd^2x' \ln \left| \vartheta_1(\frac{z'-z}{L_1}|\tau) \right|^2 \, .
\end{equation}
Up to a factor independent of $z$, this integral can be calculated by
using the quasiperiodicity \pref{jactheta} of the theta function under lattice translations:
\be{trick}
I_\xi(z) &=& \int_{a}^{a+L_1}dx' \, \int_b^{b+L_2}dy' \,\ln |\vartheta_1(\frac {z' - z} {L_1} |\tau)|^2 \nonumber \\
&=&  \int_{a-x}^{a+L_1-x}dx' \, \int_{b-y}^{b+L_2-y}dy' \, \ln |\vartheta_1(\frac {z' } {L_1} |\tau)|^2 \nonumber \\
&=& \int_{a}^{a+L_1}dx' \, \int_b^{b+L_2}dy' \,\ln |\vartheta_1(\frac {z' } {L_1} |\tau)|^2 \\
&-& \int_{a}^{a+L_1}dx' \, \int_{b-y}^{b}dy' \, \left( 4\pi \frac {y'}{L_1} + 2\pi \imag\tau  \right)  \nonumber \\
&=& I_\xi(0) +2\pi\left\{\left[ [y-(b+\half L_2)\right]^2 - (b+\half L_2)^2\right\} \nonumber  \, .
\ee
The third identity
 follows from that the integrand
is invariant under lattice translations along the $x$-axis, while 
under $z\to z+\tau$ it picks up an additional term, 
\begin{equation}
  \ln |e^{-i\pi\tau}e^{-2\pi i z/L_1} |^2 
  = 4\pi y/L_1 + 2\pi \imag\tau \, .
\end{equation}
Again taking $\xi = -(L_1+iL_2)/2$ we recover \pref{bgcon} in the text, and taking an arbitrary integration domain, 
\ie an arbitrary $\xi$ just amounts to shifting the coordinate system.

\section{Derivation of equation \pref{eq:resummed} }\label{app:resum}

We assume a compactification radius 
$R_a^2= 2p/p'$, $p,p'\in\mathbb{Z}$ --- only in this case do the vertex operators
(with integer charges) define a  rational CFT --- and
consider the following sum:
\begin{multline}\label{emsum}
 \sum_{e,m} e^{i\pi\tau \alpha^2_{e,m}} e^{-i\pi\bar\tau \bar\alpha^2_{e,m}} 
  e^{2\pi i\left[ \alpha_{e,m}Z^{(a)}-\bar\alpha_{e,m}
      \bar Z^{(a)}\right]/R_a}\\
      = \sum_{e,m} e^{i\pi\tau \alpha^2_{e,m}} e^{-i\pi\bar\tau \bar\alpha^2_{e,m}}   e^{\frac{2\pi i}{R_a^2}\left[ (e+mR_a^2/2)Z^{(a)}-(e-mR_a^2/2)\bar Z^{(a)}\right]}.
\end{multline}
We write
\begin{equation}
  mR_a^2/2 = m p/p' = (\bar m p'+r)p/p'= \bar m p + r p/p',
\end{equation}
where $r = 0,\ldots,p'-1$. Additionally, we write $e+\bar m p = 2 pk_1 +s$ and
$e-\bar m p = 2 pk_2 +s$, with $s = 0, \ldots,2p-1$. Thus, we can rewrite the
original sum \eqref{emsum} as
\begin{multline}
\sum_{r,s}\sum_{k_1} e^{i\pi \tau(2pp'k_1+s p'+r p)^2/2pp'}e^{ \frac{i\pi}{p}    (2pp'k_1+s p' +r p) Z^{(a)}}\\
\times  \sum_{k_2} e^{-i\pi\bar\tau(2pp'k_2+s p'-r p)^2/2pp'}e^{- \frac{i\pi}{p}
    (2pp'k_2+s p' -r p) \bar Z^{(a)}},
\end{multline}
or, introducing additional notation, as
\begin{equation}
  \label{eq:finitesum}
\sum_{r=0}^{p'-1}\sum_{s=0}^{2p-1} \mathcal{F}_{r,s}(Z^{(a)})\bar{\mathcal{F}}_{-r,s}(\bar Z^{(a)}),
\end{equation}
where
\be{finsum}
 {\mathcal{F}}_{r,s}(Z^{(a)})  \hskip 6cm \\ 
 = \sum_{k} e^{i\pi\tau(2pp'k+s p'+r p)^2/2pp'}e^{ \frac{i\pi}{p} 
    (2pp'k+s p' +rp) Z^{(a)}}  .  \nonumber 
\ee
This is \pref{eq:resummed}, note that Eq. \ref{eq:finitesum} is a finite sum of factorized terms.

\section{The $\nu = 3/7$  and $4/11$ states }\label{app:expstates}
As a service to the reader that does not want to penetrate the general formalism of Section \ref{wfc2}, we here provide explicit expressions for the  $\nu = 3/7$  and $4/11$  wave functions.

The latter is, as  $\nu=2/5$, a level two filling fraction, but with a quasielectron density of $1/3$. The computation of the torus wave function is in complete analogy to $2/5$ when $\mbf Q^{(2)}$ is replaced by: $\mbf Q^{(2)}=(\frac{2}{\sqrt 3}, \frac{11}{\sqrt{33}})$. The fact that now the sets $M_{\alpha}$ have different size is of no consequence for the calculation.  We find $M_1=3M_2\equiv 3M$ for $N=4M$ electrons. Using \eqref{holblock}, but with radius $R_2^2=33$ 
instead, \eqref{cb25} takes now the form:
\be{cb411}
\mathcal H_{\bar s}^{(2)}(Z_{4/11}^{(1)},Z_{4/11}^{(2)})&=&\sum_{l=0}^{2}(-1)^{tl} \mathcal G_{2j}^{1}(Z_{4/11}^{(1)},Z_{4/11}^{(2)})\nonumber\\
&&\times \mathcal G_{11l+3 s'}^{2}(Z_{4/11}^{(2)}) \, .
\ee
The integers $r_1$, $r_2$ and $t_1$, $t_2$ are fixed by \eqref{eq:pbc}. For $\phi_1=\phi_2=0$, we find $r_1=r_2=N_s-q$ and also $t_1=t_2=N_s-q$. 
The total wave function can then be written as
\be{wf411}
\Psi_{4/11} &=& \sum_{i_1<i_2<...i_{3M}\atop a_1<a_2<...a_{M}}(-1)^{\sum_j a_j} \prod_{k=1}^M \mathcal{D}_{a_k}^{(1)} \prod_{i_j<i_l}\vartheta_1(z_{i_ji_l}|\tau)^3\nonumber\\
&\times&\prod_{a_j<a_l}\vartheta_1(z_{a_ja_l}|\tau)^5\prod_{i_j, a_l}\vartheta_1(z_{i_ja_l}|\tau)^2 \nonumber\\
&\times&\mathcal{H}_{\bar s}^{(2)}(Z_{4/11}^{(1)},Z_{4/11}^{(2)}) e^{-\frac{1}{2\ell^2}\sum_k y_k^2} \, ,
\ee
where 
\be{}
Z_{4/11}^{(1)}&=& 3\sum_{j=1}^{3M} z_{i_j}/L_1+2\sum_{j=1}^{M}z_{a_j}/L_1\nonumber\\
Z_{4/11}^{(2)}&=& 11\sum_{j=1}^{M} z_{a_j}/L_1 \, .
\ee
As was the case for $\nu=2/5$, the derivatives in \eqref{wf411} are not unique and each choice yields a trial wave function. 

The  filling fraction $\nu=3/7$ is constructed by 3 vertex operators with charge vectors $\mbf Q^{(1)}=(\frac{3}{\sqrt{3}}, 0, 0)$, $\mbf Q^{(2)}=(\frac{2}{\sqrt{3}}, \frac{5}{\sqrt{15}}, 0)$ and $\mbf Q^{(3)}=(\frac{2}{\sqrt{3}}, \frac{2}{\sqrt{15}}, \frac{7}{\sqrt{35}})$. The sets are of the same size $M$, for $N=3M$ electrons. Again, we set both solenoid fluxes to zero, thus boundary conditions require $t=r=N_s-3$.  Then, \eqref{CMresult}  takes the form:
\be{cbs411}
\mathcal H_{\bar s}^{(3)}(Z_{3/7}^{(1)},Z_{3/7}^{(2)},Z_{3/7}^{(3)})&=&\sum_{l_2=0}^{2}\sum_{l_3=0}^4(-1)^{t(l_2+l_3)} \mathcal G_{2l_2+2l_3}^{1}(Z_{3/7}^{(1)})\nonumber\\
&\times& \mathcal G_{5l_2+2l_3}^{2}(Z_{3/7}^{(2)})\mathcal G_{7l_3+5 s_3'}^3(Z_{3/7}^{(3)})
\ee
with the complete wave function given by:
\be{resultwf411}
\Psi_{3/7} &=& \sum_{i_1<i_2<...i_{M}\atop {a_1<a_2<...a_{M}\atop \alpha_1,\alpha_2...\alpha_M}}(-1)^{\sigma} \prod_{k=1}^M \mathcal{D}_{a_k}^{(1)} \prod_{l=1}^M \mathcal{D}_{\alpha_l}^{(2)} \bigg[ \prod_{i_j<i_l}\vartheta_1(z_{i_ji_l}|\tau)^3\nonumber\\
&\times&\prod_{a_j<a_l}\vartheta_1(z_{a_ja_l}|\tau)^3 \prod_{\alpha_j<\alpha_l}\vartheta_1(z_{\alpha_j\alpha_l}|\tau)^3\prod_{i_j, a_l}\vartheta_1(z_{i_ja_l}|\tau)^2 \nonumber\\
&\times&\prod_{i_j, \alpha_l}\vartheta_1(z_{i_j\alpha_l}|\tau)^2 \prod_{a_j, \alpha_l}\vartheta_1(z_{a_j \alpha_l}|\tau)^2\nonumber\\
&\times& \mathcal{H}_{\bar s}^{(2)}(Z_{3/7}^{(1)},Z_{3/7}^{(2)},Z_{3/7}^{(3)})\bigg] e^{-\frac{1}{2\ell^2}\sum_k y_k^2}.
\ee
As explained earlier, $(-1)^\sigma$ is the sign picked up by rearranging the ordered fermions into sets. Antisymmetrization requires the derivatives $\mathcal{D}^{(1)}$ and $\mathcal{D}^{(2)}$ to be different.


\begin{thebibliography}{99}

{\footnotesize

\bibitem{laughlin83}R.B. Laughlin, Phys. Rev. Lett. {\bf 50}, 1395 (1983).

\bibitem{jain89}  J.K. Jain, Phys. Rev. Lett. {\bf 63}, 199 (1989).

\bibitem{multi}  See {\it e.g } S. Girvin and A.H. MacDonald, in  {\it Novel Quantum Liquids in Low-Dimensional Semiconductor Structures}, eds. S. Das Sarma and A. Pinczuk,  (Wiley, New York, 1995).

\bibitem{haldane83} F.D.M. Haldane, Phys. Rev. Lett. {\bf 51}, 605 (1983).

\bibitem{halperin84} B.I. Halperin, Phys. Rev. Lett. {\bf 52}, 1583, 2390(E) (1984). 

\bibitem{jainbook} J.K. Jain,  {\it Composite fermions}, (Cambridge University Press, 2007).

\bibitem{pan} W. Pan, {\it et al}, Phys. Rev. Lett. {\bf 90}, 016801 (2003).

\bibitem{bk2} E.J. Bergholtz, and A. Karlhede, J. Stat. Mech. (2006) L04001.

\bibitem{hans} T.H. Hansson, C.C. Chang, J.K. Jain, and S. Viefers,  Phys. Rev. Lett. {\bf 98}, 076801 (2007). 

\bibitem{CFTlong} T. H. Hansson, C.-C. Chang, J. K. Jain, and S. Viefers, Phys. Rev. B 76, 075347 (2007)

%\bibitem{natphys} E.J. Bergholtz, T.H. Hansson, M. Hermanns, and A. Karlhede, arXiv:cond-mat/0702516v1 (2007).

\bibitem{natphys} E.J. Bergholtz, T.H. Hansson, M. Hermanns, and A. Karlhede, Phys. Rev. Lett. {\bf 99}, 256803 (2007).

%\bibitem{BHHKV} E.J. Bergholtz, T.H. Hansson, M. Hermanns, A. Karlhede, and S.F. Viefers,  in preparation

\bibitem{BHHKV} E.J. Bergholtz, T.H. Hansson, M. Hermanns, A. Karlhede, and S.F. Viefers, arXiv:0712.3848 (2007).

\bibitem{wen} X.-G. Wen and Q. Niu, Phys. Rev. B {\bf 41}, 9377 (1990).

\bibitem{bk1} E.J. Bergholtz, and A. Karlhede,  Phys. Rev. Lett. {\bf 94}, 26802 (2005).

\bibitem{pfaff} E.J. Bergholtz, J. Kailasvuori, E. Wikberg, T.H. Hansson, and A.  Karlhede, Phys. Rev. B {\bf 74}, 081308 (2006).

\bibitem{seidel} A. Seidel and D.-H. Lee,   Phys. Rev. Lett. {\bf 97}, 056804 (2006).

\bibitem{read06} N. Read,  Phys. Rev. B {\bf 73}, 245334 (2006).

\bibitem{haldaneAPS}  F.D.M. Haldane, Talk at the APS March meeting (2006).

\bibitem{toruslong} E.J. Bergholtz and A. Karlhede, arXiv:0712.1927 (2007).

\bibitem{haldanewfs}  F.D.M. Haldane and E.H. Rezayi, Phys. Rev. B {\bf 31}, 2529 (1985).

\bibitem{naywil} M. Greiter, X.-G. Wen and F. Wilczek, Nucl. Phys.  B {\bf 374}, 567 (1992)

\bibitem{mooreread} G. Moore and N. Read,  Nucl. Phys. B {\bf 360}, 362 (1991).

\bibitem{yellowbook} P. DiFrancesco, P. Mathieu and D. S{\'e}n{\'e}chal,  {\it Conformal Field Theory},
(Springer, 1999).

\bibitem{haldanetorus} F.D.M. Haldane, Phys. Rev. Lett. {\bf 55}, 2095 (1985).

\bibitem{read96} N. Read and E. Rezayi,  Phys. Rev. B {\bf 54}, 16864 (1996).

\bibitem{yang} K. Yang, F.D.M. Haldane, and E.H. Rezayi, Phys. Rev. B {\bf 64}, 081301 (2001).

}

\end{thebibliography}
\end{document}